\documentclass[preprint,showpacs,preprintnumbers,amsmath,amssymb,aps]{revtex4}
\usepackage{graphicx}
\begin{document}
\title{{\bf Isospin Effects in Nuclear Multifragmentation}}

\author{ W.P. Tan$^{a}$\footnote[1]{present address: Department of Physics,
University of Notre Dame, Notre Dame, IN 46556}, S.R. Souza$^{b}$,
R.J. Charity$^{c}$, R. Donangelo$^{b}$, W.G.
Lynch$^{a}$ and M.B. Tsang$^{a}$}

\address{
$^{a}$Department of Physics and Astronomy and National Superconducting Cyclotron\\
Laboratory, Michigan State University, East Lansing, Michigan 48824\\
$^{b}$Instituto de Fisica, Universidade Federal do Rio de Janeiro\\
Cidade Universit\'aria, CP 68528, 21945-970 Rio de Janeiro, Brazil\\
$^{c}$Department of Chemistry, Washington University, St. Louis,
MO 63130}

\begin{abstract}
We develop an improved Statistical Multifragmentation Model that
provides the capability to calculate calorimetric and isotopic
observables with precision. With this new model we examine the
influence of nuclear isospin on the fragment elemental and
isotopic distributions. We show that the proposed improvements on the
model are essential for studying isospin effects in nuclear
multifragmentation. In particular, these calculations show that
accurate comparisons to experimental data require that the nuclear
masses, free energies and secondary decay must be
handled with higher precision than many current models accord.
\end{abstract}




\maketitle

\narrowtext

\section{Introduction\newline
}

\label{sec:introduction} Experiments have demonstrated that appropriately
excited nuclear systems will undergo a multifragment disintegration leading
to a final state composed of a mixture of fragments of charge $3\leq Z\leq 30
$ and light particles with $Z\leq 2$ \cite{Dasgupta01}.
Fragments are produced with large multiplicities in central heavy ion
collisions at incident energies of $E_{beam}/A\leq 100$~MeV \cite%
{bowman91,tsang93,williams97}, in larger impact parameter heavy ion
collisions at $E_{beam}/A\geq 200$~MeV \cite{ogilvie91,schuttauf96} and in
central light ion induced reactions at E$_{beam}\gtrsim $5~GeV \cite{hsi97}.
Analyses of two fragment correlations indicate breakup timescales for these
systems that are consistent with bulk disintegration \cite%
{bowman93,popescu98,Beaulieu00,Fox93,Wang99}, satisfying an important
premise of equilibrium models \cite{Gross97,Bondorf95,Souza00} that
relate multifragmentation to the nuclear liquid-gas phase transition \cite%
{Lamb78,Daniel79,Jaqaman83}.

Successful comparisons of such models have been made to the measured fragment
multiplicities, charge and energy distributions \cite%
{williams97,schuttauf96,hsi97,dagostino96}. Such success, even for
reactions where a significant collective energy of expansion is
observed \cite{williams97}, implies that these reactions populate
a significant fraction of the available phase space. Experimental
observables such as excited state and isotopic thermometers
\cite{huang,tsang97}, and the isospin dependence \cite{Xu00} of
multifragmentation, suggest a degree of thermalization less
complete for
higher incident energies or smaller systems or both \cite%
{reisdorf97,xi98,Johnston96,tsa02}. Such tests, however, have been rendered
less conclusive by the inability of many current equilibrium models to
accurately describe the later stages of the breakup where nuclear structure
details determine the spectrum of excited states and their decay branching
ratios.

Over the years, different versions of the statistical
multifragmentation models have been developed
\cite{smm,Sneppen87,Botvina87,Bondorf95,Dasgupta98}. In this paper
we based our model upon many of the theoretical foundations
described in ref. \cite{smm} and included the algorithm on
partitioning a finite system with two components as described in
ref. \cite{Sneppen87}. We call this earlier model SMM85. In
the improved Statistical Multifragmentation Model called ISMM, we
depart from the latter approach that the Helmholtz free energies
are calculated by carefully including the measured states of the
fragments, with empirical binding energies and spins
\cite{ajzenberg,Audi95,table}. We obtain expressions for these
free energies that approach the free energies of refs. \cite{smm, Sneppen87}
at excitation energies typical of excited multifragmenting
systems. The main differences between the properties of the hot
systems we calculate and those calculated in SMM85 can be
attributed to the more accurate expression for the binding
energies that we employ; the structure of the low-lying states of
the fragments plays little role in properties of the hot system.
However, these structure effects become critical when the fragments cool
later by secondary decay.

Comparisons between results from ISMM and SMM85, reveal large
differences between the predicted observables, calling many of the
previous conclusions into question. In particular, we have found
that SMM85 calculations tended to overpredict the yields of heavy
fragments, and consequently, to underestimate those of the lighter
ones. More importantly, we find that isotopic yields and
observables like the isotopic temperatures require careful
attention to the structure of the excited fragments. If such
structural effects are included many experimental trends of these
observables can be reproduced, and when they are not, the
experimental and theoretical trends are very different from each
other.

In the following, we recapitulate briefly the formalism of SMM85 and
describe in detail how we incorporate the improved structure information in
the calculation of the properties of the hot system at freezeout. This is
followed by a description of the secondary decay of the hot fragments. Then,
we turn to the comparisons of ISMM to predictions of SMM85 calculations that
take less care with these nuclear structure effects. We then compare the
present improved model to the available experimental data. Finally, we
summarize our work and provide an outlook towards future comparisons of data
to equilibrium models.

\section{The Statistical Multifragmentation Model\newline
}

\label{sec:SMM}

During the later stages of an energetic nuclear collision, the excited
system may expand to subnuclear density. This expansion may reflect the
relaxation of a compressed system formed in central collisions between
comparable mass nuclei \cite{Hsi94,Jeong94} or the thermal expansion of a
highly excited system formed in a peripheral heavy ion collision \cite%
{schuttauf96,li93,botvina95,lauret98} or in a collision between a light
projectile and a heavy nucleus \cite{beaulieu00}. For appropriate
conditions, the excited system disassembles over a time scale of 50-150 fm/c
\cite{bowman93,popescu98,Beaulieu00,Fox93,Wang99} into a mixture of
nucleons, light particles with A$\leq 4$ and heavier fragments. Equilibrium
models \cite{Gross97,Bondorf95,Souza00} such as the statistical
multifragmentation models assume that phase space is sufficiently well
occupied so that the system can be approximated by an equilibrated breakup
condition characterized by the thermal excitation energy $E^{\ast }$, the
density $\rho $, the mass $A_{0}$ and the atomic number $Z_{0}$. Then a
second \textquotedblleft freeze-out\textquotedblright\ approximation is
invoked, which assumes that the system disassembles sufficiently rapidly
that further interactions between the various particles in the equilibrated
breakup can be neglected and that subsequent secondary decay of the excited
fragments can be calculated as if these fragments are isolated.

The values of the three conserved quantities $E^{\ast },A_{0}$,
and $Z_{0}$ strongly reflect the dynamics of the excitation
process and as this dynamics lies outside SMM, they become
constraints that are introduced as input parameters to the model.
The SMM then performs the two essential tasks required of
equilibrium statistical multifragmentation models: (1) the
sampling of the equilibrium multiparticle phase space, and (2) the
secondary decay of excited fragments. The first step in sampling
the multiparticle phase space within the SMM is to select a
fragmentation mode `$m$' characterized by a set of particles
$\{N_{A,Z}\}_{m}$, which are present in the equilibrium stage. For
each fragmentation mode, mass and charge conservation provides
that:

\begin{equation}
A_0 =\sum_{\{A,Z\}} N_{A,Z}\,A \;\;\;\;\; \mathrm{and} \;\;\;\;\;
Z_0 =\sum_{\{A,Z\}} N_{A,Z}\,Z  \label{eq:azconserv}
\end{equation}

\noindent where $N_{A,Z}$ is the multiplicity of a fragment, whose mass and
atomic numbers are, respectively, $A$ and $Z$. The total multiplicity $M_{m}$
of the fragmentation mode is related to $N_{A,Z}$ by:

\begin{equation}
M_{m}=\sum_{\{A,Z\}}N_{A,Z}\;.  \label{eq:mf}
\end{equation}

\noindent The selection of the fragmentation modes and the sets of
particles $\{N_{A,Z}\}_{m}$ for each mode, is performed by an
algorithm, described in ref. \cite{Sneppen87}, that ensures that
all probable choices $ \{N_{A,Z}\}_{m}$ are sampled, but the
frequencies of sampling for the various modes do not reflect their
relative contributions to the multifragmentation phase space. This
requires the introduction of weights $ \omega_{m}$ discussed
below.

The phase space of states consistent with a decay mode
$\{N_{A,Z}\}_{m}$ reflects the number of states and consequently
the entropy consistent with that mode. The major contributions to
the total entropy are the entropies corresponding to the internal
motion, \textit{i.e.} internal excitation of the fragments. These
entropies are calculated within the SMM by introducing a
temperature $T_{m}$ for the decay mode. The ensemble average of
the expression for energy conservation can then be used to
determine the appropriate value of $T_{m}$ as follows:

\begin{equation}
E_{A_{0},Z_{0}}^{\mathrm{G.S.}}+E^{*}=C_{C}\frac{Z_{0}^{2}}{A_{0}^{1/3}}%
\left( \frac{V_{0}}{V_{m0}}\right)
^{1/3}+\sum_{\{A,Z\}}N_{A,Z}E_{A,Z}(T_{m})\;. \label{eq:econserv}
\end{equation}

\noindent On the left hand side of the equation, the total energy
is decomposed into the total ground state and total excitation
energies of the source. The ground state energy,
$E_{A_{0},Z_{0}}^{\mathrm{G.S.}}$, represents the ground state
energy of the source calculated as a single spherical nucleus.
The first term on the right is the
Coulomb energy of a homogeneous sphere of matter containing the
total charge $Z_{0}$ and mass $A_{0}$, which is
evaluated at a density $\rho=\rho _{0}\left( V_{0}/V_{m0}\right) $ where $%
\rho_{0}$ is the saturation density and $V_{0}$ and $V_{m0}$ are
the volumes occupied by the system at saturation and at the
breakup densities, respectively. The remaining terms on the right
hand side are energy contributions, \textit{i.e.} the kinetic,
ground state, extra Coulomb, and excitation energies of the
individual fragments that are specified below.
For the Coulomb energy, this decomposition is enabled by invoking
a modified Wigner Seitz approximation \cite{wigner}, whose
accuracy for the multifragmentation process has been explored in
refs. \cite{smm,Sneppen87}. The result of
applying Eq.(\ref{eq:econserv}) is to obtain values for $T_{m}$
which conserve energy for the ensemble averaged mean for each
decay mode and consequently fluctuate from one decay mode to
another reflecting the corresponding variations in the Coulomb,
kinetic and ground state energies of the collection of fragments
that characterize each decay mode.

The weight $\omega_{m}$ for each decay mode is calculated by evaluating the
corresponding number of states for the mode

\begin{equation}
\omega_{m}=\exp(S_{m}),\;
\end{equation}

\noindent where the total entropy $S_{m}$ of the mode is obtained by summing
the contributions from each particle

\begin{equation}
S_{m}=\sum S_{A,Z},\;
\end{equation}

\noindent where both $E_{A,Z}$ and $S_{A,Z}$ are obtained from the Helmholtz
free energies $F_{A,Z}$ via the usual thermodynamical relations:

\begin{equation}
S=-\frac{\partial F}{\partial T}
\end{equation}

\noindent and

\begin{equation}
F=E-TS\;,
\end{equation}

\noindent which apply to both the contributions from individual fragments
and to their overall sums of $E_{m},S_{m}$ and $F_{m}$.

The contributions $F_{A,Z}(T_{m})$, associated with each fragment in the
partition may be decomposed into four terms:

\begin{eqnarray}
F_{A,Z}(T) &=& E^{\mathrm{G.S.}}_{A,Z}-C_C\frac{Z^2}{A^{1/3}} \left(\frac{V_0%
} {V_{m0}}\right)^{1/3}  \nonumber \\
&+& F_{A,Z}^{K}(T)+F^*_{A,Z}(T)\;,  \label{eq:fazterms}
\end{eqnarray}

\noindent where $E^{G.S.}_{A,Z}$ is the ground state energy of the fragment.
The kinetic term corresponds to:

\begin{eqnarray}
F_{A,Z}^{K}(T) &=&-T\ln\left[g_{A,Z}V_{f0}\left[\frac{m_NAT}{2\pi\hbar^2} %
\right] ^{3/2} \right]  \nonumber \\
&+& T\ln\left(N_{A,Z}!\right)/N_{A,Z}\;.  \label{eq:fekin}
\end{eqnarray}

\noindent In this expression, $V_{f0}=V_{m0}-V_{0}$ is the free volume, $%
m_{N}$ represents the nucleon mass, and $g_{A,Z}$ is the spin degeneracy
factor. Empirical ground state spin degeneracy factors are used for $A<5$
because these nuclei have no low lying excited states. For simplicity, we
take $g_{AZ}=1$ for heavier nuclei because the influence of non-zero spins
on $F_{A,Z}^{K}(T)$ is small and can be compensated by small changes in the
level density expression for the fragment. The Coulomb term, $-C_{C}\frac{%
Z^{2}}{A^{1/3}}\left( \frac{V_{0}}{V_{m0}}\right) ^{1/3}$ in Eq.~(\ref%
{eq:fazterms}) represents the corrections in the Wigner-Seitz approximation
for the individual particles. The excitation of the intrinsic degrees of
freedom is taken into account by $F_{A,Z}^{\ast}(T)$, and is zero for light
particles with no excited states.

To calculate the properties of the multifragment emission from the excited
source, one should sum the contributions of all the partitions consistent
with energy, mass and charge conservation. Such a procedure, however, would
be extremely time consuming owing to the huge number of possible modes.
Therefore, the present approach samples the more probable modes via a Monte
Carlo calculation. This is discussed in detail in ref.~\cite{Sneppen87}; we
note in passing that the Monte Carlo procedure introduces a bias since not
all the mass and charge partitions enter with the same weight. Therefore $%
\omega_{m}$ must be modified to correct for this bias \cite{Sneppen87}.

Taking these modifications into account, the average value of a physical
observable $O$ is calculated by taking a weighted average,

\begin{equation}
\langle O\rangle=\frac{\sum_{m}\omega_{m}O_{m}}{\sum_{m}\omega_{m}}\;.
\label{eq:avevalue}
\end{equation}
This average applies both to observables calculated from the primary
distributions and from the secondary distributions. Because the weights are
not unity, the calculation of the statistical uncertainties associated with
the Monte Carlo procedure requires care. They can be easily obtained,
however, by repeating the Monte Carlo procedure with a different
initialization of the random number generator and calculating the variance
of the fluctuations in the predicted observables.

\subsection{\noindent\noindent Ground state energies}

\label{sec:gse} Since the predicted primary yields of excited
fragments are exponentially related to their binding energies
\cite{Bondorf95}, it is natural to assume that accurate values for
the ground state masses for the observed fragments are needed. In
addition, the isospin dependence of the masses and consequent
yields of heavier nuclei away from the valley of stability can
influence the predicted yields of measured light nuclei closer to
the valley of stability because all yields must be consistent with
the constraints imposed by mass and charge conservation. To
provide more accurate predictions of isotopic distributions, it is
relevant to
replace the somewhat inaccurate Liquid Drop Mass (LDM) parameterization \cite%
{Sneppen87,smm} used by many current SMM codes \cite{Tan01,tsang01}.

\begin{figure}[tbp]
\includegraphics[width=11cm]{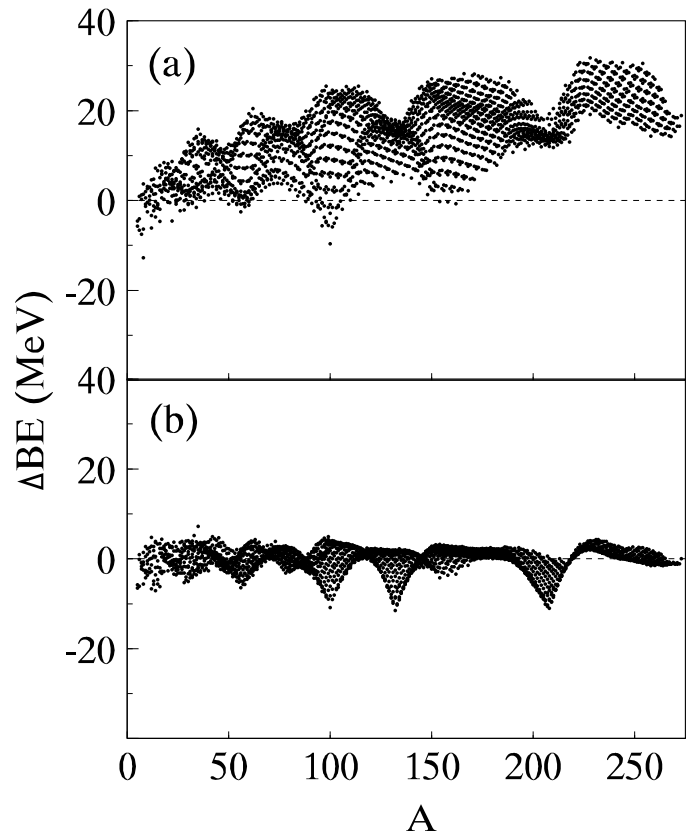}
\caption{Differences between the {\it total} binding energies
predicted by mass formulae and those recomended in ref.\
\protect\cite{Audi95}. Upper panel :Plot (a) displays the
differences when one uses the LDM mass formula used in
SMM85 calculations. \protect\cite {Sneppen87}. Lower panel:
Plot (b) displays the differences when one uses the ILDM mass
formula presented in this work.} \label{fig:diffbe}
\end{figure}

To address this problem, we use the recommended binding
energies values from Audi and Wasptra ~\cite{Audi95} when
available. The sampling of the most probable partitions discards
extremely exotic fragments, which would contribute with a
vanishing statistical weight. Nonetheless, applications of the SMM
to realistic multifragmentation scenarios require the code to
predict the binding energies for many nuclei that have not been
measured. Therefore, we use a more accurate description of
unknown masses given in ref.~\cite{Preston}:

\begin{eqnarray}
B_{A,Z}^{ILDM} &=&C_{V}A-C_{S}A^{2/3}-C_{C}\frac{Z^{2}}{A^{1/3}}  \nonumber
\\
&+&\delta _{A,Z}A^{-1/2}+C_{d}\frac{Z^{2}}{A}\;,  \label{eq:ld}
\end{eqnarray}

\noindent where

\begin{equation}
C_{i}=a_{i}\left[ 1-k\left( \frac{A-2Z}{A}\right) ^{2}\right]
\label{eq:ldmav}
\end{equation}%
\noindent and $i={V},{S}$ stand for volume and surface, respectively. The
coefficient $\delta _{A,Z}$ corresponds to the usual pairing term:
\begin{equation}
\delta _{A,Z}=
\begin{cases}
+\delta_{pairing} & \text{$N$ and $Z$ even}\\
0 & \text{$A$ odd}\\
-\delta_{pairing} & \text{$N$ and $Z$ odd}\\
\end{cases}
\label{eq:pairing}
\end{equation}%
The parameters corresponding to the best fit of the empirical
masses in ref.~ \cite{Audi95} are $a_{V}$=15.6658 MeV, $a_{S}$ =
18.9952 MeV, $k$=1.77441, $C_{C}$ = 0.720531 MeV,
$\delta_{pairing}$ = 10.8567 MeV and $C_{d}$ = 1.74859 MeV. To
illustrate the improvement in the model, the upper panel (a) of
Figure \ref{fig:diffbe} shows the difference between the
calculated binding energies from the parameterization of the LDM
of ref.~ \cite{Sneppen87} used in most current SMM codes and the
empirical values. The lower panel (b) shows the corresponding
comparison between the calculated binding energies using
Eq.~(\ref{eq:ld}) with the improved parameters (ILDM) and the
empirical values. One should note that the \textit{ total} binding
energies are plotted, rather than the binding energy per nucleon.
This improved agreement suggests that the predictions for
unmeasured masses will also be improved.

\begin{figure}[tbp]
\includegraphics[width=11cm]{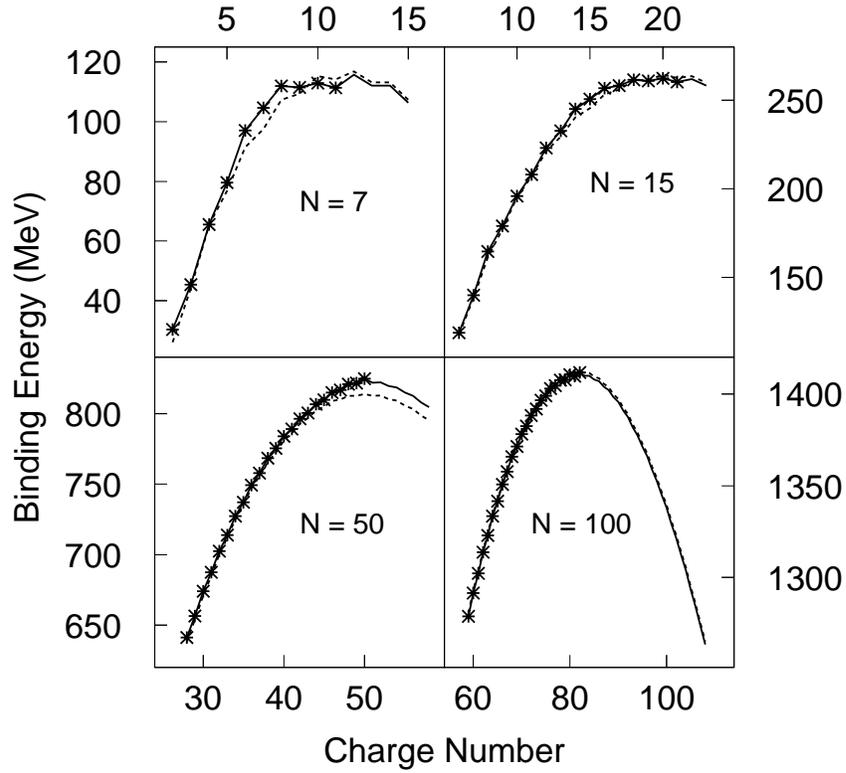}
\caption{Total binding energies for various nuclei. The full lines
correspond to the corrected LDM formula, whereas the symbols
represent the experimental data of ref.\ \protect\cite{Audi95}.
The dashed lines correspond to the predictions given by Eq.\
(\protect\ref{eq:ld}) using the optimized parameters. For details
see the text.} \label{fig:extrap}
\end{figure}

Despite the improvement in the overall mass predictions, there can be
discontinuity between the extrapolated (dashed line) and empirical values
(points) as illustrated in Fig. \ref{fig:extrap}. To improve the matching
between the binding energies of the known masses and the ones predicted by
our mass formula, we compute average shifts of the ILDM formula from the
empirical values and use these shifts to correct the values in Eq.\ (\ref%
{eq:ld}). For an isotone that has a lower charge than its isotonic partners
in the compilation of ref.~\cite{Audi95} we use the three lightest isotones
with the same value of $N$ in the compilation to compute the shift.
Similarly for an isotone that has a higher charge than its isotonic partners
in the compilation of ref.~\cite{Audi95} we use the three heaviest isotones
in the compilation to compute the shift. This shift is then subtracted from
the prediction of the ILDM formula:

\begin{equation}
B_{A,Z}^{extrap}=B_{A,Z}^{ILDM}-\Delta _{N}\;,  \label{eq:extrap}
\end{equation}

\noindent where

\begin{equation}
\Delta_{N}=\frac{1}{3}\sum\limits_{i=1}^{3}\left(
B_{A_{i},Z_{i}}^{ILDM}-B_{A_{i},Z_{i}}^{Audi}\right) ,
\label{eq:beshift}
\end{equation}

\noindent $B_{A_{i},Z_{i}}^{Audi}$ is the corresponding value from
the compilation of ref.~\cite{Audi95}. Two shift values are
therefore computed for each value of $N$. The final binding energy
values used in the ISMM calculations are illustrated for four
cases by the solid lines in Fig.~\ref{fig:extrap} where it shows
that the discontinuity between the empirical (star) and
extrapolated (dashed line) values is removed.

\subsection{\noindent Fragment internal free energies}

\label{sec:ife}

In this work, we have modified SMM85 so as to allow accurate predictions of
isotopic properties, but have limited the extent of these modifications in
an effort to retain many of the predictions of the original theory. In
particular, we have retained the high temperature properties of the fragment
free energies, $F_{A,Z}^{\ast }$, which are parameterized here and in the
SMM85 as:

\begin{equation}
F^{*}_{A,Z}(T)=\beta_{0}A^{2/3}\left[ \left( \frac{T_{C}^{2}-T^{2}}{%
T_{C}^{2}+T^{2}} \right) ^{5/4} -1\right] -A\frac{T^{2}}{\epsilon_{0}}\;,
\label{eq:intfree}
\end{equation}

\noindent where $\beta _{0}=18.0$ MeV, $\epsilon _{0}=16.0$ MeV, and $%
T_{C}=18.0$ MeV. This expression holds only for temperatures
smaller than critical temperature, $T_{C}$. At low temperatures,
$T<<T_{C}$, this expression depends quadratically on $T$ as
expected for a Fermi liquid. At the critical temperature where the
surface tension vanishes, the surface energy contribution to the
total free energy $F_{A,Z}(T)$ falls to zero when the surface
energy contribution in Eq. (\ref{eq:intfree}) is combined with the
corresponding ground state energy term in Eq.~(\ref{eq:fazterms}).
As we do not calculate decays at $T>10$ MeV, we do not concern
ourselves here with the form for $F_{A,Z}^{\ast }(T)$ at $T\geq
T_{C}$. For 3~MeV~$\lesssim T\lesssim $~10~MeV, where
multifragmentation is important, however, this form for
$F_{A,Z}^{\ast }(T)$ in Eq. (\ref{eq:intfree}) is not unique, and
other expressions with different thermal properties should be
explored. In the following we introduce empirical modifications to
this free energy expression by taking into account the nuclear
structural information of known excited states.

First we turn our attention to the fact that most fragments at
$T>2$~MeV are particle unstable and will sequentially decay after
freezeout. This decay is sensitive to nuclear structure properties
of the excited fragments such as their nuclear levels, binding
energies, spins, parities and decay branching ratios. The first
three of these quantities also influence the free energies; this
can be calculated via the fragment internal partition functions.
Self-consistency in the freeze-out approximation dictates that the
states from which these fragments decay after freezeout should be
consistent with the Helmholtz free energies used in calculating
the primary yields of the hot fragments at freeze-out.

In order to discuss this self-consistency requirement, we must consider the
density of states $\rho_{states}(E)$ and its mathematical relationship with
the Helmholtz free excitation energy $F^{*}(T)$:

\begin{equation}
e^{-F^{*}/T}=\int_{0}^{\infty}\,dE\,e^{-E/T}\rho_{states}(E),
\label{eq:ferho}
\end{equation}

\noindent where the integral is over the excitation energy $E$ of the
nucleus. Here we have, for simplicity, neglected the complications of a
degenerate ground state, which contributes negligibly to the free energy at
high excitation energy. In the original papers on the SMM, the level
densities corresponding to the SMM were not stipulated. We now consider what
is required of the density of states to achieve the high temperature
behavior for $F^{*}_{A,Z}(T)$ given by Eq.~(\ref{eq:intfree}). Then we will
address the general issue of making the level densities consistent with
empirical information and how that impacts the free energies. Finally, we
will discuss specific details of the incorporation of the empirical
information into the level density expressions.

\subsubsection{High temperature behavior}

First we investigate what forms of level densities may be consistent with
the free energies in Eq.~(\ref{eq:intfree}). We note that the functional
dependence of $F_{A,Z}^{\ast }(T)$ used in Eq.(\ref{eq:intfree}) makes its
analytical inversion difficult at high temperatures. Instead, it is easier to
find a smooth real functional form for $\rho _{states}(E)$ that reproduces
the numerical values for $F_{A,Z}^{\ast }(T)$ at high temperatures than it
would be to perform an inverse Laplace transformation of $F_{A,Z}^{\ast }(T)$
in the complex plane. We note that if one inverts a Taylor expansion of $%
F_{A,Z}^{\ast }(T)$ up to second order in $T$ by the saddle point
approximation, one obtains the Fermi gas expression:

\begin{equation}
\rho_{FG,states}(E)=\frac{a_{SMM}^{1/4}}{\sqrt{4\pi}E^{3/4}} \exp\left( 2%
\sqrt{a_{SMM}E}\right)  \label{eq:rhoFg}
\end{equation}

\noindent where $a_{SMM}$ is the absolute value of the coefficient of the
second order term of the free energy expansion in $T$:

\begin{equation}
a_{SMM} = \frac{A}{\epsilon_{0}}+\frac{5}{2}\beta_{0}\frac{A^{2/3}}{T_{c}^{2}%
}.  \label{eq:aSMM}
\end{equation}

\begin{figure}[tbp]
\includegraphics[width=11cm]{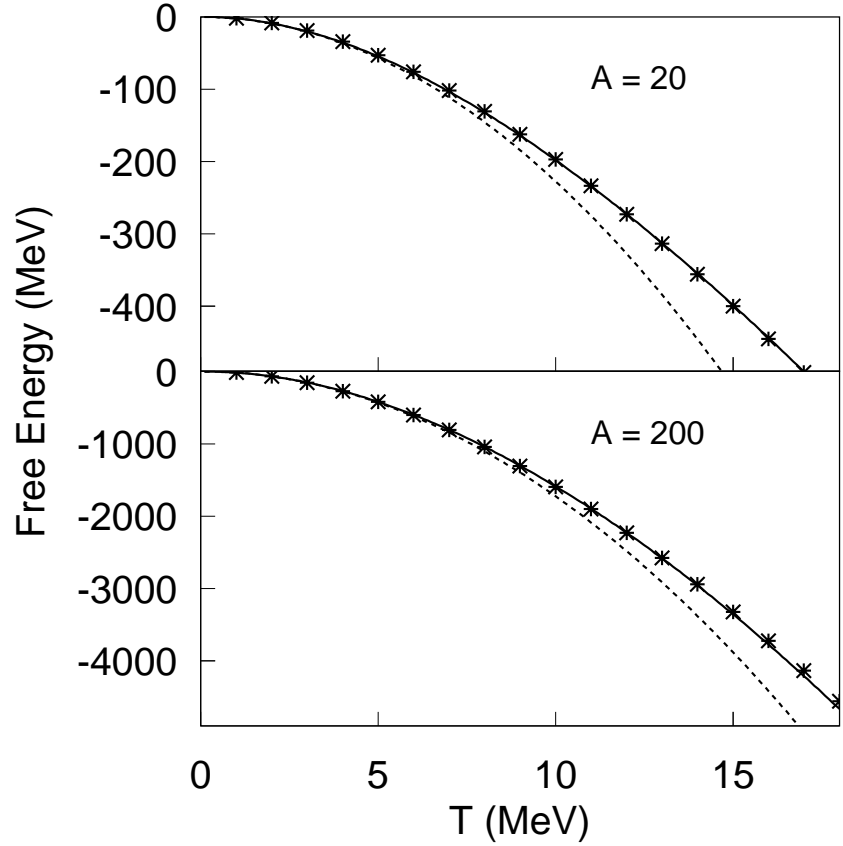}
\caption{Internal free energies for $A=20$ (upper panel) and
$A=200$ (lower panel). The SMM85 expression [Eq.\
(\protect\ref{eq:intfree})] is represented by the full line
whereas the dashed lines stand for the results obtained with the
Taylor expansion [Eq.\ (\protect\ref{eq:rhoFg})]. The Free energy
calculated through the level density given by Eq.\ (\protect\ref
{eq:rhocor}) is depicted by the symbols.} \label{fig:feSMM}
\end{figure}

\noindent However, this expression is unsatisfactory at high temperatures,
as is illustrated in Fig.~\ref{fig:feSMM} when the free energies obtained
from Eq. \ref{eq:rhoFg} (dashed lines) are compared with SMM85 free energies
in Eq. \ref{eq:intfree} (solid lines). Instead, we take Eq. \ref{eq:rhoFg}
as a starting point and obtain a useful analytic expression by multiplying $%
\rho _{FG,states}(E)$ by an \textit{ad hoc} energy dependent term to obtain
free energy values in numerical agreement with Eq.\ (\ref{eq:intfree}):

\begin{equation}
\rho_{SMM,states}(E)=\rho_{FG,states}(E)\,e^{-b_{SMM}(a_{SMM}E)^{3/2}},
\label{eq:rhocor}
\end{equation}

\noindent where $b_{SMM}$ is given by:

\begin{eqnarray}  \label{eq:tauSMM}
b_{SMM} &=& 0.07 A^{-\tau}, \\
\tau &=& 1.82 \left(1+\frac{A}{4500}\right).
\end{eqnarray}

\noindent The free energies obtained via Eqs.\ (\ref{eq:ferho}) and (\ref%
{eq:rhocor}) are displayed in Fig.~\ref{fig:feSMM} as symbols for two
different mass regions. This simple parameterization is fairly accurate at
temperatures $T\leq10$ MeV, in the range of interest.

\subsubsection{Empirical Level densities at low excitation energies for Z$
\leq$15}

\label{sect:ldzleq15}

Several factors motivate the efforts to develop an accurate
treatment for the level densities at low excitation energy for
$Z\leq 15$. The first factor is that most multifragmentation data are
available for light fragments in this mass range. The second is
that empirical nuclear structure information is also available for
these nuclei. A comparable treatment of the level density for the
heavier fragments would be interesting, but the needed structure
information is frequently incomplete or entirely missing.
Fortunately, if we focus on the yields for $A\leq 8$, the
contributions from the secondary decay of the heavy nuclei with $Z
> 15$ are of the order of $10$\%. Thus the errors introduced by
the neglect of this structure information for the heavy nuclei
does not strongly influence the results of the final yields and
one can proceed towards reasonable predictions at the present
time.

\begin{figure}[tbp]
\includegraphics[width=11cm]{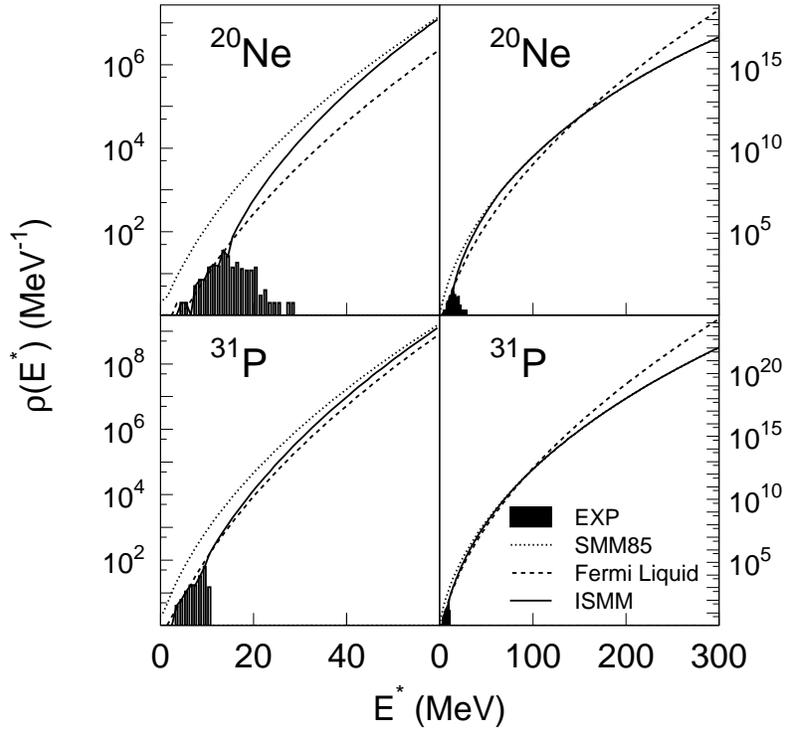}
\caption{Level densities as a function of excitation energy for
$^{20}$Ne and $^{31}$P. Two energy ranges are plotted to show the
behaviors of level densities at both low (left panels) and
high energy (right panels) ends. The
density of experimentally known levels is shown as bars in the low
energy region. The dashed lines are the extrapolations of the
empirical values according to Eq. \protect\ref{eq_rhocon}. The
dotted lines are the level density (Eq.  \protect \ref{eq:rhocor})
parametrized from the SMM85. The solid lines are the level
density adopted in this work (Eqs.
\protect\ref{eq_rho1}-\protect\ref {eq_rho3}).}
\label{fig:density}
\end{figure}

At lower excitation energies, it is customary to discuss the density of
levels $\rho _{levels}$ rather than the density of states because this
definition is more useful experimentally when the spins of specific levels
are not accurately known. Mathematically, the density of states is related
to the densities of levels for individual spin values $\rho _{levels}(E,J)$
by:
\begin{equation}
\rho _{states}(E)=\sum\limits_{J}\left( 2J+1\right) \rho _{levels}(E,J)
\end{equation}%
While the spacings between energy levels in a given nucleus generally
decrease smoothly with excitation energy, as a practical matter one often
decomposes the empirical level density $\rho _{emp,levels}(E,J)$ into two
expressions that apply in two different approximate excitation energy
domains: (1) one (labelled as $\rho _{D,levels}(E,J)$) containing discrete
well separated states at low excitation energies and (2) another (labelled
as $\rho _{C,levels}(E,J)$) containing a continuum of overlapping states at
higher excitation energies. For Z$\leq $15, empirical level information \cite%
{ajzenberg,table} is applied as much as possible to the low-lying discrete
level density, wherever the experimental level scheme seems complete,
\begin{equation}
\rho _{D,levels}(E,J)=\sum_{i}\delta (E_{i}-E),  \label{eq_rhoexp}
\end{equation}%
where the summation runs over the excitation energies $E_{i}$ corresponding
to states of spin $J$. Examples of empirical levels for $^{20}$Ne and $^{31}$%
P are shown as bars in Fig.\ \ref{fig:density}. For higher excitation
energies, a good approximation to the continuum level density has been
obtained by ref. \cite{Gilbert65} by combining Fermi liquid theory, a simple
spin dependence and experimental knowledge. The relevant expressions, shown
as dashed lines in Fig.\ \ref{fig:density}, are \cite{Chen88},
\begin{equation}
\rho _{C,levels}(E,J)=\rho _{C}(E)f(J,\sigma )\text{ for }E>E_{c}
\label{eq_rhocon}
\end{equation}%
where
\begin{eqnarray}
\rho _{C}(E) &=&\frac{\exp [2\sqrt{a(E-E_{0})}]}{12\sqrt{2}%
a^{1/4}(E-E_{0})^{5/4}\sigma }, \\
f(J,\sigma ) &=&\frac{(2J+1)\exp [-(J+1/2)^{2}/2\sigma ^{2}]}{2\sigma ^{2}},
\\
\sigma ^{2} &=&0.0888\sqrt{a(E-E_{0})}A^{2/3},
\end{eqnarray}%
and the level density parameter $a=A/8$. $E^{\ast }$, $J$, A and Z are the
excitation energy, spin, mass and charge numbers of the fragment. $E_{0}$ is
determined by matching the total high-lying level density to the total
low-lying level density as follows,
\begin{equation}
\int_{E_{0}}^{E_{c}}dE\int dJ\rho _{C,levels}(E,J)=\int_{0}^{E_{c}}dE\int
dJ\rho _{D,levels}(E,J),  \label{eq_rhomatch}
\end{equation}%
where $E_{c}$ is the energy at which the switch from discrete to continuum
level density expressions is made.

The comparison in Eq. (\ref{eq_rhomatch}) is between the total level
densities summed over spin. This is done primarily to reduce the sensitivity
in the matching to uncertainties in the spin assignments for some of the
discrete states. By adjusting the parameter $E_{0}$, the total level density
for continuum states was connected smoothly to the total level density for
low-lying states at $E<E_{c}$ and $Z<12$. The connection point $E_{c}$ to
high-lying states, for $Z<12$, was chosen to be the maximum excitation
energy up to which information concerning the number and locations of
discrete states appears to be complete so that the empirical level density
(Eq.\ref{eq_rhoexp}) was solely applied for low-lying states.

For the case of $Z\geq12$, low-lying states are not well identified
experimentally and a continuum approximation to the discrete level density
\cite{Chen88} was used by modifying the empirical interpolation formula of
Ref. \cite{Gilbert65} to include a spin dependence:
\begin{eqnarray}
\rho _{D,levels}(E,J) &=&\frac{1}{T_{1}}\exp [(E-E_{1})/T_{1}]  \nonumber \\
&\times& \frac{(2J+1)\exp [-(J+1/2)^{2}/2\sigma _{0}^{2}]}{
\sum_{i}(2J_{i}+1)\exp [-(J_{i}+1/2)^{2}/2\sigma _{0i}^{2}]},
\label{eq_rholow}
\end{eqnarray}
for $E\leq E_{c}$, where the spin cutoff parameter $\sigma_{0}^{2}=0.0888%
\sqrt{a(E_{c}-E_{0})}A^{2/3}$. For $Z\geq12$, the values of $%
E_{c}=E_{c}(A,Z) $ were taken from Ref. \cite{Gilbert65} as well
as parameters $T_{1}=T_{1}(A,Z)$ and $E_{1}=E_{1}(A,Z)$, and in
this case, the approximate level density (Eq.\ref{eq_rholow}) was
used in place of an empirical level density for the low-lying
states.

\subsubsection{Matching low and high excitation energy behavior}

Now, we turn to the requirement of self consistency between the expression
for $F^{*}_{A,Z}(T)$ and the level density relevant to secondary decay. In
general, secondary decay becomes more sensitive to nuclear structure
quantities such as the excitation energies, spins, etc. as the systems
decay towards the ground state. At low excitation energies, one is more
accurate using empirical level densities in place of the expression in Eq.\
( \ref{eq:intfree}), which does not even depend on Z. As the excitation
energy is increased, however, the continuum level density becomes very
large, little sensitivity to nuclear structure details remains and a simpler
expression like Eq.\ (\ref{eq:intfree}) may suffice.

In the following, we take $\rho _{SMM,states}(E)$ to be the state
density at high energies and match it to the continuum part of the
empirical state densities at low excitation energies. This
procedure uses the empirical information for excitation energies
$E^{\ast }<E_{c}$, a linear interpolation for $E_{c}<E^{\ast
}<E_{c}+\Delta E$, and $\rho _{SMM,states}(E)$ at higher values of
the excitation energy. The net result is a set of level density
and state density expressions that span the range
of excitation energies relevant to multifragmentation phenomena. For $%
E^{\ast }<E_{c}$, one uses the expression for the discrete, low-lying state
density,
\begin{equation}
\rho _{ISMM}(E,J)=\rho _{D}(E^{\ast },J).  \label{eq_rho1}
\end{equation}%
For $E_{c}<E^{\ast }<E_{c}+\Delta E$, the new level density is an
interpolation involving the continuum expression relevant at low
excitation energies between $\rho _{C,states}$ and $\rho
_{SMM,states}$,
\begin{eqnarray}
\rho _{ISMM}(E^{\ast },J) &=&\rho _{C}(E^{\ast },J)(1-\frac{E^{\ast }-E_{c}}{%
\Delta E})  \nonumber \\
&&+\rho _{SMM}(E^{\ast },J)\frac{E^{\ast }-E_{c}}{\Delta E},  \label{eq_rho2}
\end{eqnarray}%
where $\Delta E=2.5A$ MeV provides a smooth transition from $\rho _{C}$ to $%
\rho _{SMM}$. The SMM level density (shown as dotted lines in Fig.\ \ref%
{fig:density}) can be incorporated with a similar spin dependence as in Eq. %
\ref{eq_rhocon},
\begin{equation}
\rho _{SMM}(E^{\ast },J)=\rho _{SMM}(E^{\ast })f(J,\sigma ).
\label{eq_rhosmmj}
\end{equation}%
For $E^{\ast }>E_{c}+\Delta E$, the new density simply becomes the same as
the SMM level density $\rho _{SMM}$,
\begin{equation}
\rho _{ISMM}(E^{\ast },J)=\rho _{SMM}(E^{\ast },J).  \label{eq_rho3}
\end{equation}

\noindent In Fig.\ \ref{fig:density}, the empirically modified level density
described in Eqs. (\ref{eq_rho1}-\ref{eq_rho3}) is plotted as solid lines
for $^{20}$Ne and $^{31}$P.

The level density $\rho _{C}$ in Eq. (\ref{eq_rhocon}) can be used as a
proper extension to the low-lying level density $\rho _{D}$ in Eqs. (\ref%
{eq_rhoexp}) and (\ref{eq_rholow}) and a bridge for matching to the SMM
level density at continuum. Such a matching procedure provides a state
density that is empirically based at low excitation energies but becomes
progressively more uncertain as the excitation energy is increased above $%
E^{\ast }\approx E_{c}.$ This uncertainty in the thermal properties of
nuclei at such high excitation energies is not a question of finding an
appropriate interpolation, but is, in fact, a fundamental issue that must be
resolved by comparisons to experimental data. For example, other expressions
can be proposed for the level density at $E^{\ast }>E_{c}$ and comparisons
of experimental data to SMM predictions of sensitive multifragment
observables can be used to constrain the level densities at high excitation
energies.

\begin{figure}[tbp]
\includegraphics[width=11cm]{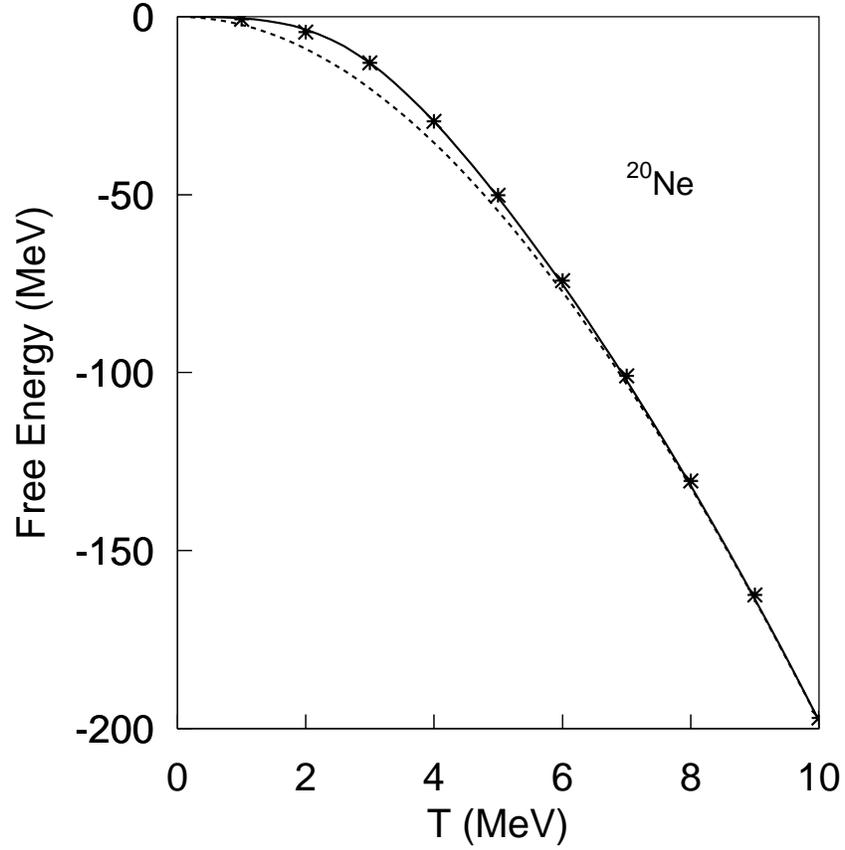}
\caption{Comparison between $F^*(T)$ calculated through Eqs.\
(\protect\ref {eq:ferho}) and
(\protect\ref{eq_rho1})-(\protect\ref{eq_rho3}), symbols, and the
approximation given by Eq.\ (\protect\ref{eq:fintfit}), full line.
To illustrate the influence of quantum effects at low
temperatures, the dashed line represents the free energy used in
SMM85 calculations Eq.\ (\protect\ref{eq:intfree}). For
details see text.} \label{fig:fefit}
\end{figure}

Free energies $F^{*}_{A,Z}(T)$, which reflect contributions from the
discrete excited states are obtained by inserting this parameterization for $%
\rho_{states}(E)$ into Eq.\ (\ref{eq:ferho}), and performing a numerical
integration. To facilitate the insertion of these free energies into the SMM
algorithm, we parameterize $F^{*}_{A,Z}(T)$ by:

\begin{equation}
F^{*}_{A,Z}(T)=F^{*}_{SMM}(T)\left( 1-\frac{1}{1+\exp[(T-T_{adj})/\Delta T]}%
\right) \;,  \label{eq:fintfit}
\end{equation}

\noindent where $F_{SMM}^{\ast }(T)$ stands for the SMM internal
free energy of Eq.\ (\ref{eq:intfree}) which is adopted in various
SMM models. The parameters $T_{adj}$ and $\Delta T$ are adjusted
to reproduce the numerical calculation of $F^{\ast }(T)$ provided
by Eqs.\ (\ref{eq:ferho}) and (\ref{eq_rho1}-\ref{eq_rho3}) for
$T\leq 10$~MeV. In these fits, a value for $\Delta T=1.0$~MeV is
used for most nuclei (The exceptions are mainly very light
nuclei.), while $T_{adj}$ is varied freely. The accuracy of the
fit is illustrated in Fig.\ \ref{fig:fefit}, which compares the
exact values of $F^{\ast }(T)$ (symbols) to the approximation
given by Eq.\ (\ref{eq:fintfit}) (solid line), for a $^{20}$Ne
nucleus. The dashed line in this figure represents the free energy
used in SMM85 calculations in which the experimental discrete
levels are neglected. The matching procedure allows the discrete
excited states to dominate the low temperature behavior, while the
high temperature behavior remains similar to that of the SMM85,
consistent with the goals stated above.

\begin{figure}[tbp]
\includegraphics[width=11cm]{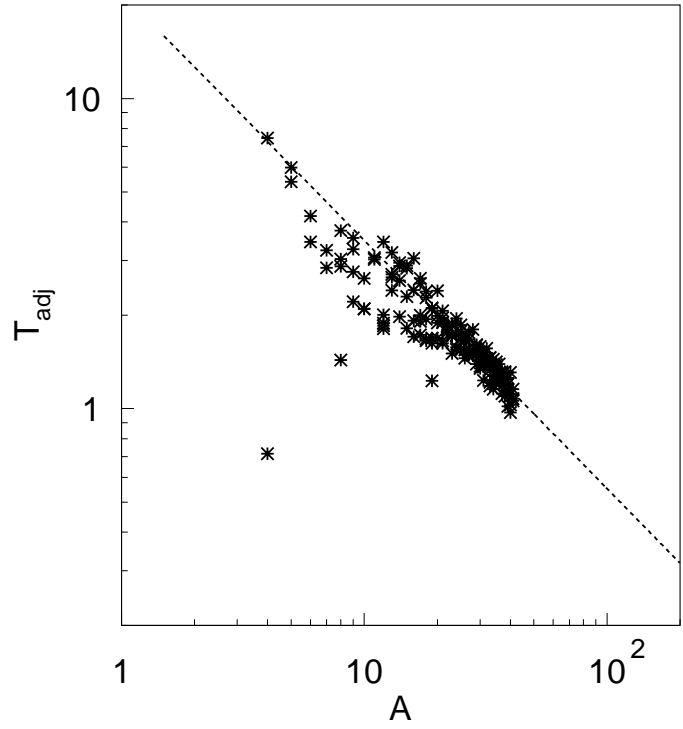}
\caption{Best
fit values of $T_0$ for different nuclei (symbols). The dashed
line corresponds to Eq. \ref{eq:extrapt0} used for $Z > 15$.}
\label{fig:t0}
\end{figure}

Because the empirical level densities vary from nucleus to nucleus, the
parameters $T_{adj}$ and $\Delta T$ must be fitted for each nucleus used to
obtain $F_{A,Z}^{\ast }(T)$. Fits of the same quality as that for $^{20}$Ne
are achieved for all the light nuclei with $Z\leq 15$. These fitted values
of $T_{adj}$ are shown as symbols in Fig. \ref{fig:t0}. We do not perform
such fits for $Z>15,$ because the level density information there is less
complete. We nevertheless extrapolate the main trend of the parameters to
heavy nuclei, for which detailed experimental information on discrete
excited states is not available, in order to avoid spurious discontinuities
in the equilibrium primary yields. As mentioned above, there seems to be a
very weak dependence on $\Delta T$ and, therefore, we assume $\Delta T=1.0$%
~MeV for $Z>15$. In spite of the uncertainty in extrapolating $T_{adj}$, the
dashed line in Fig.\ \ref{fig:t0} shows that

\begin{equation}
T_{adj}=22.0A^{-0.8}\; \mathrm{MeV}\;\; (Z>15)  \label{eq:extrapt0}
\end{equation}

\noindent describes the trend (dashed line) for the lower masses and we
adopted it for the higher masses as well.

\section{Secondary Decay\newline
}

\label{sec:secdecay}

With few exceptions, the stable yields after secondary decay are
the quantities that are usually measured experimentally. An
accurate secondary decay procedure is indispensable to calculate
the contributions from secondary decay and deduce the information
of the primary hot system from experimental data. The sequential
decay procedure consists of two parts. One is to decay particles
with Z$\leq $15 through a large empirical (MSU-DECAY) table
including all the states of nuclei with known information such as
binding energy, spin, isospin, parity and decay branching ratios.
The other part is to use the Gemini code \cite{charity} for
particles outside the empirical table (usually Z$>$15).

\subsection{Decay table}

\label{sect:deczleq15}

The implementation of Eqs. (\ref{eq_rho1}-\ref{eq_rho3}) involves the
construction of a 'table' of quantities such as the excitation energies,
spins, isospins, and parities of levels of nuclei with $Z\leq 15$. For
excitation energies $E<E_{c}$ and $Z\leq 15$, each of the entries in the
table corresponds to one of the tabulated empirical levels. When the
information on the level is complete, it is used. For known levels with
incomplete spectroscopic information, values for the spin, isospin, and
parity were chosen randomly as follows: spins of 0-4 (1/2-9/2) were assumed
with equal probability for even-A (odd-A) nuclei, parities were assumed to
be odd or even with equal probability, and isospins were assumed to be the
same as the isospin of the ground state. This simple assumption turns out to
be sufficient since most of spectroscopic information is known for these
low-lying states.

For excitation energies where little or no structure information
exists, the level density was assumed to be given by the level
density algorithm discussed in the
previous section and groups of levels were binned together in discrete excitation
energy intervals of 1 MeV for $%
E^{\ast}<15$ MeV, 2 MeV for $15<E^{\ast}<30$ MeV, and 3 MeV for
$E^{\ast}>30$ MeV in order to reduce the computer memory
requirements. The results of the calculations do not appear to be
sensitive to these binning widths. A cutoff energy of
$E_{cutoff}^{\ast}/A=5$ MeV was introduced corresponding to a mean
lifetime of the continuum states at the cutoff energy about 125
fm/c. For simplicity, parities of these states were chosen to be
positive and negative with equal probability and isospins were
taken to be equal to the isospin of the ground state of the same
nucleus.

\subsection{Sequential decay algorithm}

Before sequential decay starts, hot fragments from primary breakup need to
be populated over the sampled levels in the prepared table according to the
temperature. For the $i$th level of a given nucleus (A,Z) with its energy $%
E^{*}_{i}$ and spin $J_{i}$, the initial population is,
\begin{equation}  \label{eq_inipop}
Y_{i} = Y_{0}(A,Z)\frac{(2J_{i}+1)\exp(-E^{*}_{i}/T)\rho(E^{*}_{i},J_{i})} {
\sum_{i}(2J_{i}+1)\exp(-E^{*}_{i}/T)\rho(E^{*}_{i},J_{i})}
\end{equation}
where $Y_{0}$ is the primary yield of nucleus (A,Z) and T is the temperature
associated with the intrinsic excitation of the fragmenting system at
breakup.

Finally all the fragments will decay sequentially through various
excited states of lighter nuclei down to the ground states of the
daughter decay products. Eight decay branches of n, 2n, p, 2p, d,
t, $^{3}$He and alpha were considered for the particle unstable
decays of nuclei with Z$\leq $ 15. The decays of particle stable
excited states via gamma rays were also taken into account for the
sequential decay process and for the calculation of the final
ground state yields. If known, tabulated branching ratios were
used to describe the decay of particle unstable states. Where such
information was not available, the
branching ratios were calculated from the Hauser-Feshbach formula \cite%
{Hauser52},

\begin{equation}  \label{eq_branch}
\frac{\Gamma_{c}}{\Gamma} = \frac{G_{c}}{\sum_{d} G_{d}}
\end{equation}
where
\begin{eqnarray}  \label{eq_tcoeff}
G_d &=& \langle I_d I_e I_{d3} I_{e3}|I_p I_{p3}\rangle^2  \nonumber \\
&& \times \sum^{|J_d+J_e|}_{J=|J_d-J_e|} \sum^{|J_p+J|}_{l=|J_p-J|} \frac{
1+\pi_p \pi_d\pi_e (-1)^l}{2}T_l(E)
\end{eqnarray}
for a given decay channel $d$ (or a given state of the daughter fragment). $%
J_{p} $, $J_{d}$, and $J_{e}$ are the spins of the parent, daughter and
emitted nuclei; $J$ and $l$ are the spin and orbital angular momentum of the
decay channel; $T_{l}(E)$ is the transmission coefficient for the $l$th
partial wave. The factor $[1+\pi_{p}\pi_{d}\pi_{e} (-1)^{l}]/2$ enforces
parity conservation and depends on the parities $\pi=\pm1$ of the parent,
daughter and emitted nuclei. The Clebsch-Gordon coefficient involving $I_{p}$%
, $I_{d}$, and $I_{e}$, the isospins of the parent, daughter and emitted
nuclei, likewise allows one to take isospin conservation into account.

For decays from empirical discrete states and $l\leq20$, the transmission
coefficients were interpolated from a set of calculated optical model
transmission coefficients; otherwise a parameterization described in Ref.
\cite{Chen88} was applied.

\section{Model Predictions and Comparisons}

\subsection{Caloric Curve}

\begin{figure}[tbp]
\includegraphics[width=11cm]{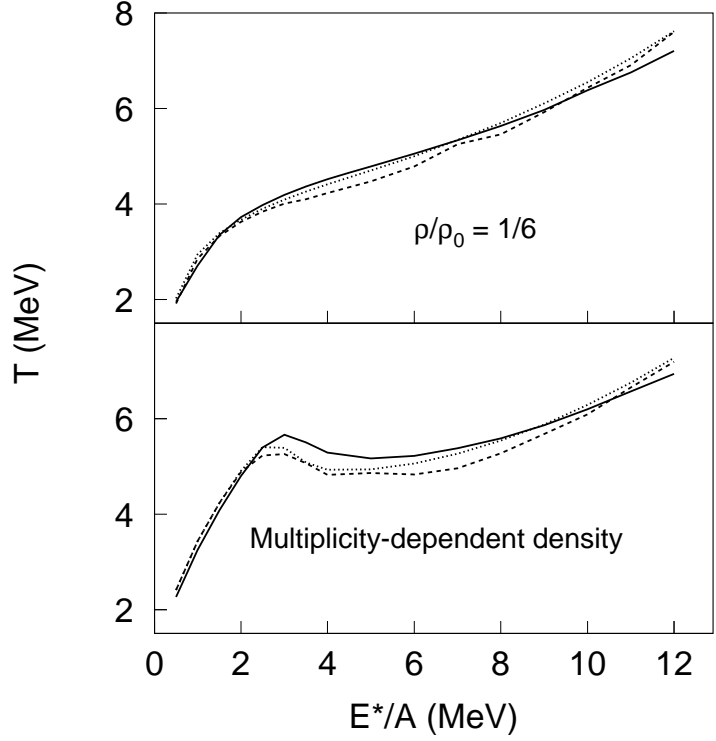}
\caption{Caloric curves are shown for calculations of the system
of A=168 and Z=75 at fixed breakup density and
multiplicity-dependent density. The dotted lines are calculated
from the SMM85. The dashed lines result when empirical binding
energies are taken into account. The solid lines are obtained from
the improved model, ISMM, with empirical modifications of both
binding energies and free energies.} \label{fig:caloric}
\end{figure}

Before presenting predictions for isotope distributions and other
observables for which the present theoretical developments were
undertaken, we examine predictions of the present improved model
for the caloric curve and the primary fragment multiplicities,
both of which displayed features in SMM85 and other SMM
calculations \cite{williams97} that are characteristic of low
density phase transition. For example, SMM85 calculations predict
an enhanced heat capacity for multifragmenting systems reflecting
the latent heat for transforming nuclear fragments (Fermi liquid)
into nucleonic gas. Fig.\ref{fig:caloric} shows the caloric curve,
i.e. the dependence of the mean fragmentation temperature
$\left\langle T_{m}\right\rangle $ on excitation energy, for a
system with A$_{0}$=168 and Z$_{0}$=75. In both panels, the
dotted lines indicate the relationships predicted by the original SMM85 \cite%
{smm,Sneppen87}, the solid lines indicate the corresponding
predictions of the ISMM with all the modifications discussed in
this paper and the dashed lines indicate the results provided by
an SMM85 calculation that uses the new binding energies of Eqs.
(\ref{eq:ld}-\ref{eq:beshift}) and the old parameterization of
ref. \cite{smm} for the Helmholtz free energies. These latter
calculations allow one to assess the impact of the changes in the
binding energies and free energies independently.

\begin{figure}[tbp]
\includegraphics[width=11cm]{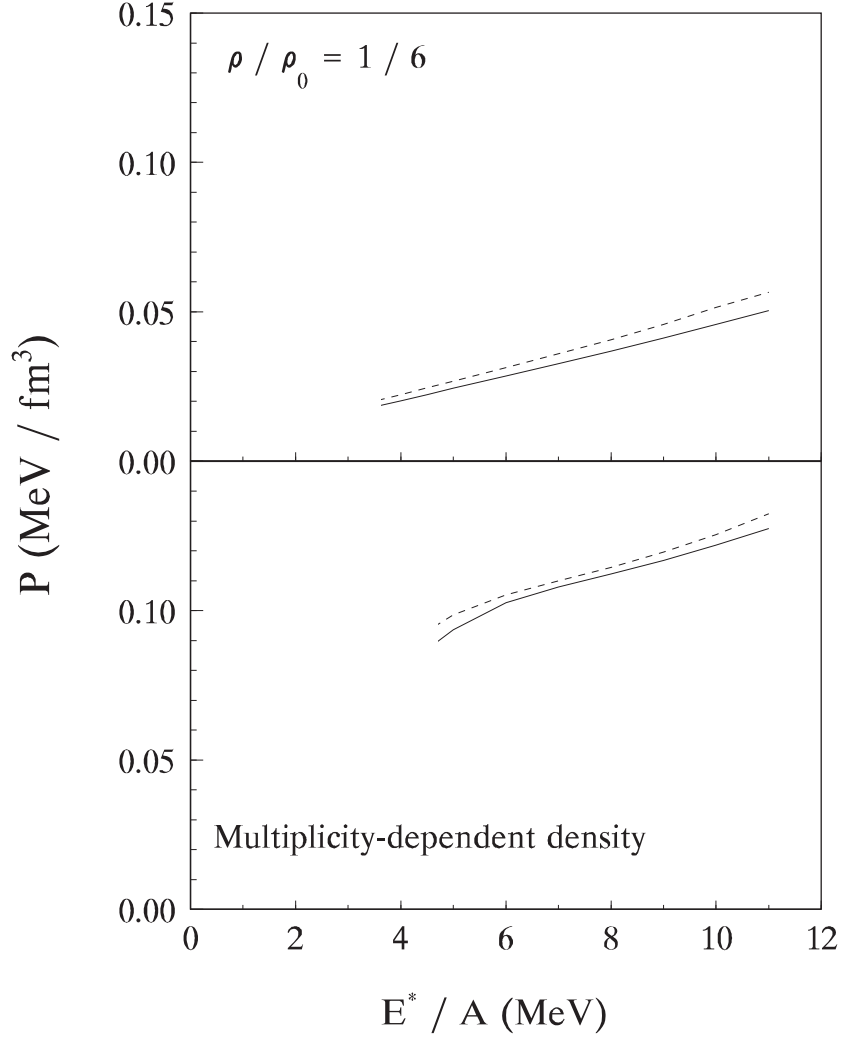}
\caption{Pressure curves due to kinetic motion and Coulomb
interaction (see Eq. \ref{eq:pressure}) are plotted for the system
of A=168 and Z=75 at fixed breakup density and
multiplicity-dependent density. The dotted lines are calculated
from the SMM85 while the ISMM presents the solid
lines.} \label{fig:pressure}
\end{figure}

\begin{figure}[tbp]
\includegraphics[width=11cm]{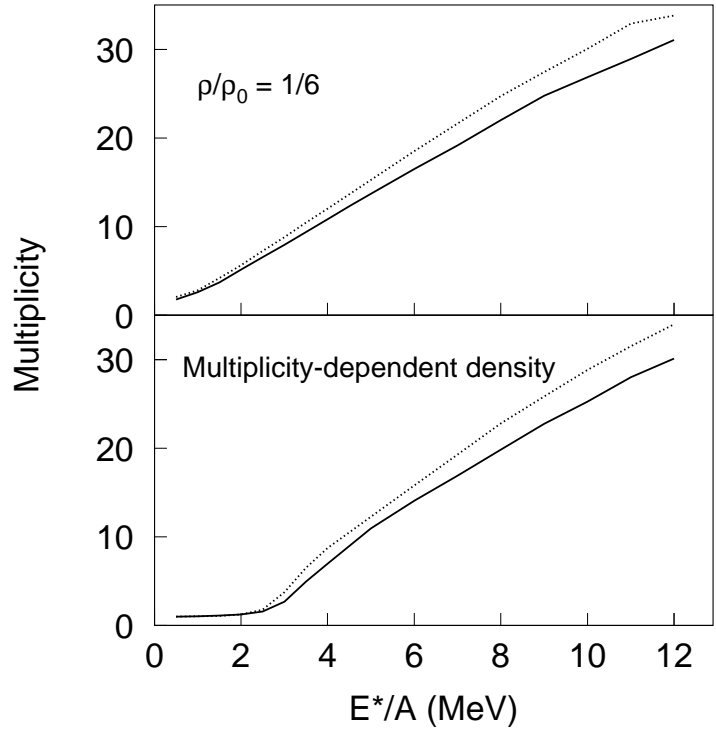}
\caption{Average breakup multiplicities are shown for the system
of A=168 and Z=75 at fixed breakup density and
multiplicity-dependent density. The dotted lines are calculated
from the SMM85 while the ISMM presents the solid
lines.} \label{fig:multiplicity}
\end{figure}

The two panels provide the caloric curves corresponding to two
different constraints on the density. In the lower panel, a
multiplicity-dependent breakup density \cite{smm} is assumed,
corresponding to a fixed interfragment spacing at breakup; this
leads to a pronounced plateau in the caloric curve for all three
calculations. By taking into account the kinetic motion and the
Coulomb interaction, we have estimated the pressure using the
relationship

\begin{equation}
P=[\frac{\partial F}{\partial V_{m0}}]_{T,N_{A,Z}}\approx\frac{(M-1)\cdot T}{%
V_{f}}+\frac{Z_{0}^{2}\ e^{2}}{5RV},  \label{eq:pressure}
\end{equation}

\noindent where $P$ is the pressure, $M$ is the total
multiplicity, $V_{f}$ is the free breakup volume and $V$ is the
total volume. Limiting the pressure estimates to temperatures for
which the multiplicity exceeds 10 and the pressure can be more
reasonably defined, we show the pressure corresponding to these
multiplicity-dependent breakup densities in the lower panel of
Fig. \ref{fig:pressure}. The corresponding primary
fragment multiplicities are shown in the lower panel of Fig. \ref%
{fig:multiplicity}. Consistent with the conclusions of ref.
\cite{dasgupta}, we find the requirement of approximately constant
interfragment spacing corresponds to breakup pressures that
exhibit only a small fractional increase with temperature. In the
upper panels, we show the corresponding caloric curves
(Fig.\ref{fig:caloric}), pressures (Fig. \ref{fig:pressure}), and
multiplicities (Fig. \ref{fig:multiplicity}) calculated at fixed
breakup density $\rho/\rho_{0}=1/6$. These show a steeper
dependence of the caloric curves on excitation energy and the
small maximum displayed in the lower panel of
Fig.\ref{fig:caloric} at excitation energies of about 3 MeV
disappears. The corresponding pressures at constant density, shown
in the upper panel of Fig. \ref{fig:pressure}, increase
monotonically with excitation energy. However, they are lower than
those calculated assuming a multiplicity dependent breakup
density, because the density for the constant volume calculations
is lower.

These figures reveal that the trends of the thermal dynamical
properties of these three models to be similar. In general, the
temperatures in the plateau region at $E^{\ast }/A=3-8$~MeV in the
lower panel of Fig. 7 are larger for the ISMM calculations using
the improved free excitation energies. This is consistent with the
fact that the level densities and, consequently, the entropies of
the fragments are lower in the improved model, which generally
raises the temperature corresponding to a given excitation energy.
Specifically in the plateau region, reducing the entropies of the
fragments raises the latent heat for the transformation from
excited fragments to nucleon gas and raises the temperature at
which the transition occurs. The influence of the improved binding
energies on the caloric curve is less obvious, but this change
seems to be largely responsible for the differences between the
SMM85 and ISMM at $E^{\ast }/A>6$~MeV.

\begin{figure}[tbp]
\includegraphics[width=11cm]{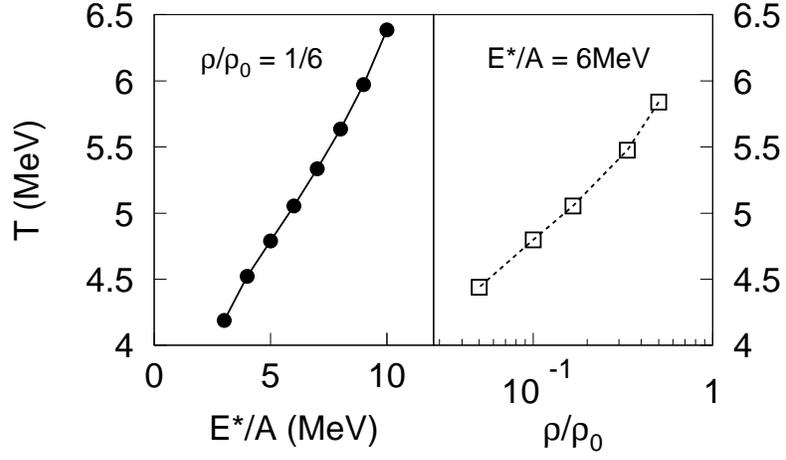}
\caption{Dependences of temperature on excitation energy and
breakup density are shown for the system of A=168 and Z=75.
Calculations as function of excitation energy at fixed density of
1/6 normal density are shown as solid circles in the left panel.
Calculations as
function of density at fixed excitation energy are shown as open
squares in the right panel.} \label{fig:terho}
\end{figure}

Discussions of the nuclear caloric curve usually focus on the
excitation energy dependence of the temperature and ignore the
density dependence. To illustrate that the phase diagram is two
dimensional and a density dependence does exist, we contrast in
Fig. \ref{fig:terho} the density dependence (right panel) of the
temperature at a fixed excitation energy of $ E^{\ast }/A=6MeV$
(open squares) to the excitation energy dependence (left panel) of
the temperature at a fixed density of $\rho /\rho _{0}=1/6$ (solid
circles). Both the excitation energy and the density dependences
of the caloric curve are clearly important. It is therefore
relevant to find and measure observables that constrain
significantly the freezeout density.

\subsection{Charge and Mass Distributions}

\begin{figure}[tbp]
\includegraphics[width=11cm]{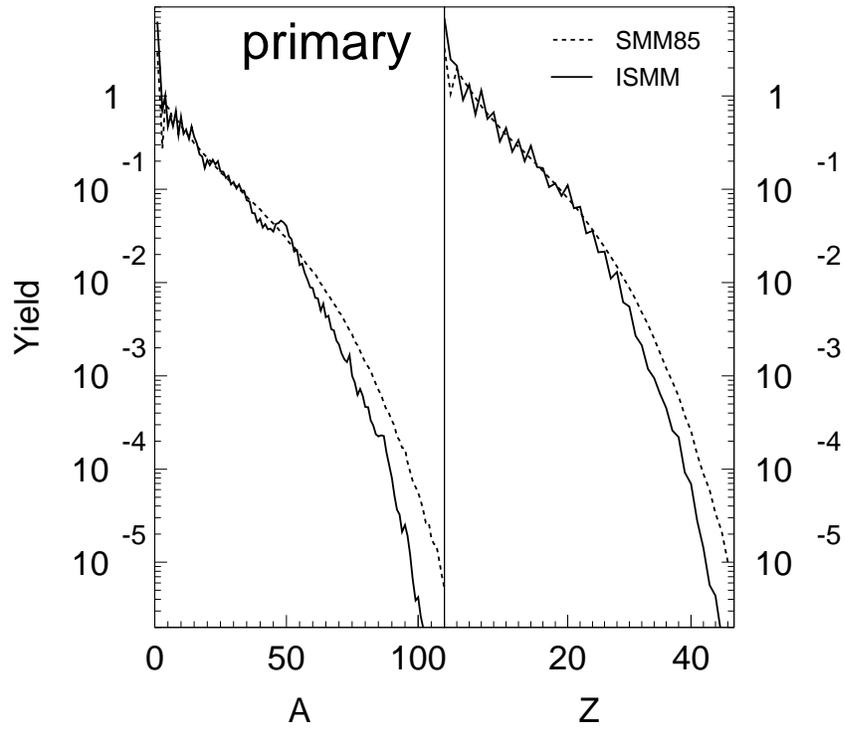}
\caption{Mass and charge distributions for the system of
A=186 and Z=75. The dashed lines are the calculations from the
SMM85. The solid lines are calculated using the improved
model ISMM.} \label{fig:az}
\end{figure}

Calculations of the mass distribution (left panel) and charge distribution
(right panel) for excited primary fragments are shown in Fig. \ref{fig:az}
for a system with $A_{0}=186$ and $Z_{0}=75$ at $E^{\ast}/A=6MeV$. This
system has the same charge to mass ratio as the symmetric $^{124}$Sn+$^{124}$%
Sn system, but is chosen to be 3/4 of the total mass in order to
approximately address the mass loss to preequilibrium emission.
The dotted lines denote the predictions using SMM85 and the solid
lines denote the predictions using ISMM. The primary distributions
from ISMM fluctuate about the smooth distributions of SMM85 for
$Z<20$ and $A<60$ and then fall below SMM85 at higher mass and
charge. The fluctuations are related to the influence of shell and
pairing effects on the ground state masses, which have no
significant impact on the final yields after secondary decay as
discussed below. The trend of reduced yields
at higher masses and charges is related to the tendency shown in Fig. \ref%
{fig:diffbe} for the binding energies in the SMM85 to consistently
exceed the empirical values at $Z>20$ and $A>60$. Because
conservation of mass and charge dictates that an increase in the
yields of heavier fragments must be compensated by a decrease in
the yields of the lighter ones, one should see a comparable
under-prediction of the primary yields of the lighter fragments by
SMM85.

\begin{figure}[tbp]
\includegraphics[width=11cm]{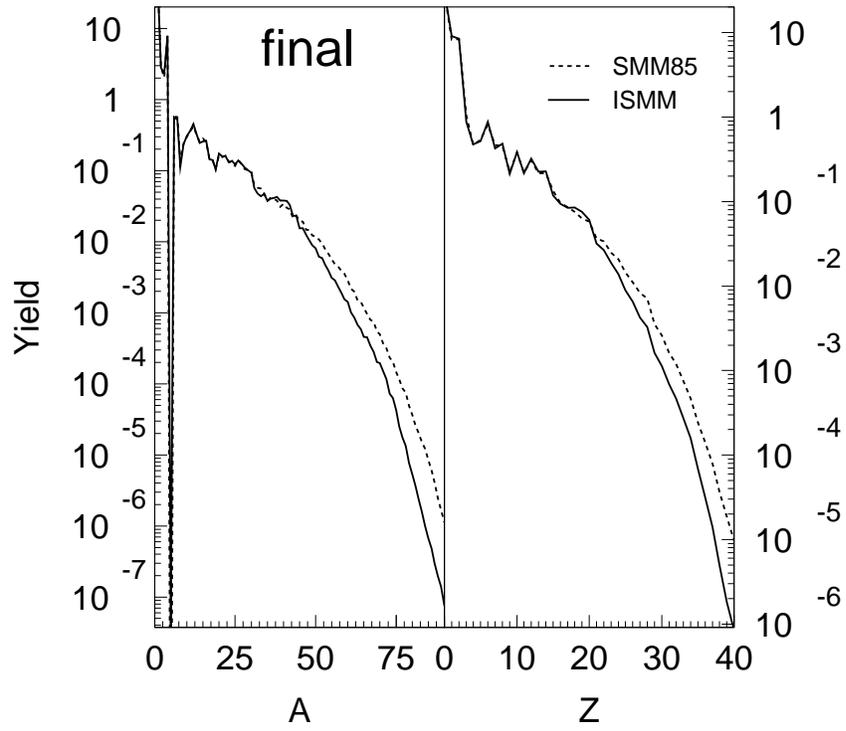}
\caption{Final mass and charge distributions after applying MSU-DECAY, the
empirical secondary decay procedure discussed in Sect.
\protect\ref{sec:secdecay}. The dashed lines are calculated from
the primary results of the SMM85 while the solid lines are
from ISMM.} \label{fig:azsmooth}
\end{figure}

\begin{figure}[tbp]
\includegraphics[width=11cm]{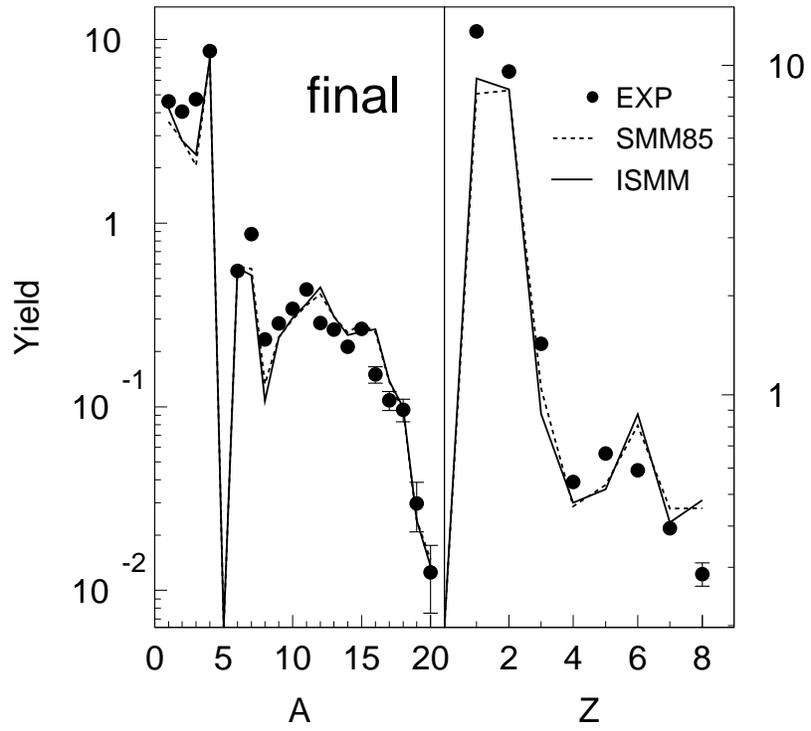}
\caption{Final mass and charge distributions from
ISMM (solid lines) and SMM85 (dashed lines) are shown. For reference, some
measured data from refs. \cite{Xu00,txliu} are plotted as solid
circles.} \label{fig:azexp}
\end{figure}

Fig. \ref{fig:azsmooth} shows the corresponding final mass (left
panel) and charge (right panel) distributions after secondary
decay. The solid lines denote the predictions using ISMM.
Experimental fragment yields from the central $^{124}Sn+^{124}Sn$
collisions are plotted as solid points \cite{txliu}. To
investigate the influence of the fluctuations in the primary
distributions due to shell and pairing effects on the ground state
masses, we have decayed the primary fragments from the SMM85 via
the same empirical secondary decay procedure discussed in Sect.
\ref{sec:secdecay}. The final mass and charge distributions of the
SMM85 are shown as the dashed lines in Fig. \ref{fig:azexp}.
Minimal discrepancies are seen in low mass and charge regime
indicating that the secondary decay mechanism washes out the
fluctuations in the primary distributions due to the influence of
shell and pairing effects on the ground state masses. Meanwhile,
significant differences on heavy fragments remain. In order to see
the differences between the two calculations, the low A and Z
regions are expanded in Figure \ref{fig:azexp}. Here again,
experimental fragment yields from the central $^{124}Sn+^{124}Sn$
collisions are plotted as solid points \cite{txliu}. The agreement
is very good even though no special attempt has been made to
optimize the parameters of the calculations to achieve the best
representation of the data.

\subsection{Isotopic distributions}

\begin{figure}[tbp]
\includegraphics[width=11cm]{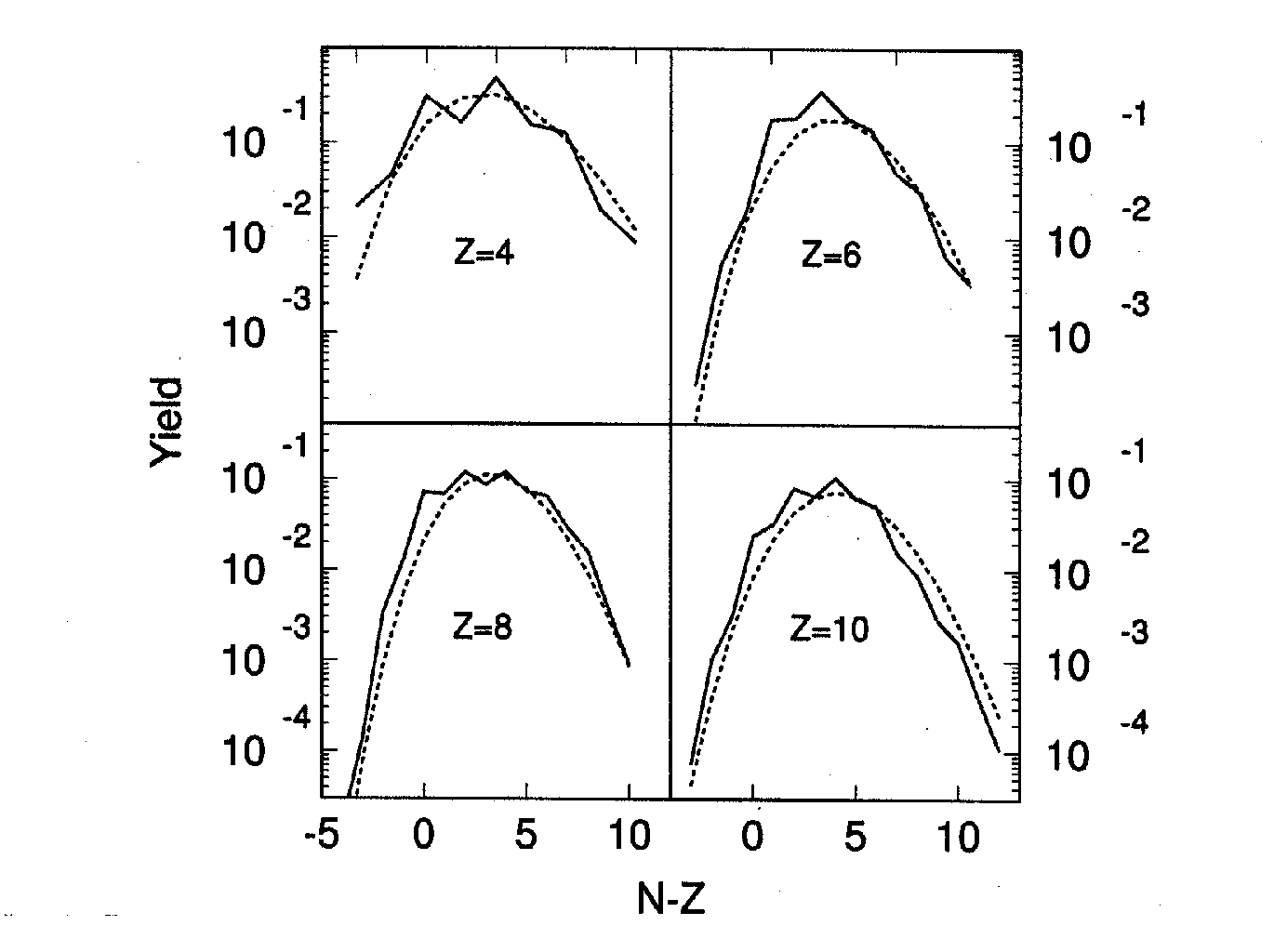}
\caption{Primary isotopic distributions for Be, C, O and
Ne nuclei. The dashed lines correspond to the calculations of the
SMM85 while the solid lines
represent the results of ISMM.} 
\label{fig:priiso}
\end{figure}

\begin{figure}[tbp]
\includegraphics[width=11cm]{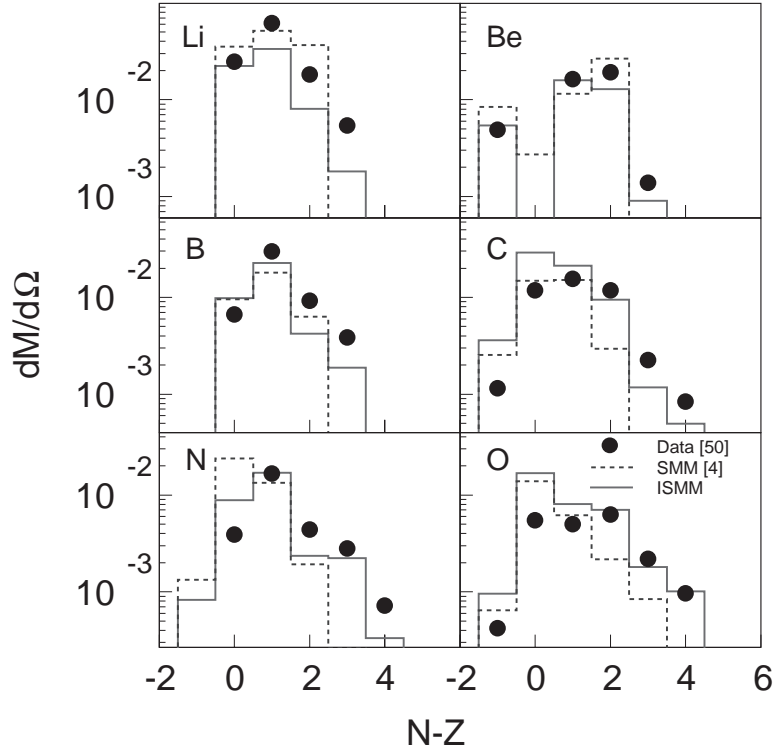}
\caption{Isotopic distributions are shown for isotopes from Li to
O. Experimental data are shown as the solid circles. The dashed
lines denote calculations from the SMM code used in
ref.\cite{williams97} and the solid lines are the final
distributions obtained using the present ISMM model, which
contains an empirical secondary decay procedure.}
\label{fig:finiso}
\end{figure}

In Fig. \ref{fig:priiso}, the primary isotopic distributions for
four elements emitted are shown for a system with $A_{0}=186$ and
$Z_{0}=75$ at $E^{\ast}/A = 6$ MeV. The solid lines show
predictions for the present improved model and the dashed lines
show predictions of the SMM85 code of refs. \cite{smm,Sneppen87}.
The two calculations produce primary isotopic distributions that
are considerably broader and more neutron rich than corresponding
final distributions after secondary decay shown in Fig.
\ref{fig:finiso}. For reference, the measured isotopic
distributions of refs. \cite{Xu00,txliu} are shown as solid points
in Fig. \ref{fig:finiso}. While the parameters of the code were
not optimized to reproduce the data, it is interesting to note
that the widths of the distributions from ISMM calculations and
data are similar although the data seem to be more neutron rich
than the calculations. Studies have shown that the final isotopic
distributions calculated with an empirical secondary decay
procedure such as that employed by the ISMM are much broader and
more neutron-rich than the corresponding distributions predicted
by the more schematic statistical models \cite{Souza00}. In order
to compare with the available experimental data, the isospin
observables derived from these isotopic distributions such as
isoscaling parameters \cite {Xu00,tsang01,Tan01} and isotopic
temperatures require an accurate secondary decay approach with
detailed nuclear structure information taken into account.

Isotope thermometers have been utilized as the primary probes for
extracting the caloric curve of the nuclear liquid-gas phase
transition. Since these observables are constructed from the
isotopic distributions, they share the sensitivity to structure
effects in the secondary decay discussed above. In the isotopic
thermometer technique, the temperature is extracted from a set of
four isotopes produced in multifragment breakups as follows
\cite{Albergo} ,
\begin{equation}
T_{iso}=\frac{\Delta B}{\ln (aR)}  \label{eq_albergo}
\end{equation}
where
\begin{eqnarray}
R
&=&\frac{Y(A_{1},Z_{1})/Y(A_{1}+1,Z_{1})}{Y(A_{2},Z_{2})/Y(A_{2}+1.Z_{2})},
\label{eq_tratio}\\
\Delta B &=&B(A_{1},Z_{1})-B(A_{1}+1,Z_{1})  \nonumber \\
&&-B(A_{2},Z_{2})+B(A_{2}+1,Z_{2}),
\end{eqnarray}
and
\begin{equation}
a=\frac{\left( 2J_{Z2,A2}+1\right) \left(
2J_{Z1,A1+1}+1\right) }{\left( 2J_{Z1,A1}+1\right) \left(
2J_{Z2,A2+1}+1\right) }\left[ \frac{A_{2}\left( A_{1}+1\right)
}{A_{1}\left( A_{2}+1\right) }\right] ^{3/2}.
\end{equation}
Here $Y(A,Z)$ is the yield of a given fragment with mass A and charge Z; $%
B(A,Z)$ is the binding energy of this fragment; and $J_{Z,A}$ is
the ground state spin of the nucleus. Although this expression is
derived within the context of the grand canonical ensemble, it has
been applied to a wide variety of reactions and regarded as an
effective or \textquotedblleft apparent\textquotedblright\
temperature that may differ somewhat from the true freezeout
temperature $T$. The relationship between $T_{iso}$ and $T$ can be
calculated within an appropriate statistical model for the
fragmentation process if one exists. In general, one chooses
a set of four isotopes with large $\Delta B$ to minimize
sensitivity to details of the corrections from secondary decay.

To examine the influence of secondary decay, measured and
calculated temperatures are extracted from double ratios of Z=2-8
fragments and plotted in Figure \ref{fig:tlio}. The large $\Delta
B$ requirement generally limits the apparent temperature
observables to three types of thermometers: a.) $
T_{iso}(^{3,4}He)$, $Z_{2}$=2, $A_{2}$=3, b.)
$T_{iso}(^{11,12}C)$, $Z_{2}$ =6, $A_{2}$=11, and c.)
$T_{iso}(^{15,16}O)$, $Z_{2}$=8, $A_{2}$=15, where thermometer (a)
involves the light particle pair $^{3,4}$He while thermometers (b) and (c)
concern only the intermediate mass fragments (IMF's) of Z=3-8.
Table I lists the corresponding thermometers plotted in Figure
 \ref{fig:tlio} . The
top left panel in Fig. \ref{fig:tlio} shows the ISMM predictions
for these three types of thermometers as a function of $A_{1}$.

Since the denominator in Eq. (\ref{eq_tratio}) is fixed by
classifying the temperatures into three types, the fluctuations
are related to $A_{1}$. In all cases, the two thermometers
involving $^{10}$Be and $^{18}$O are much higher than the others
due to many low lying states in these nuclei \cite{tsang97}. The
extracted temperatures from all the other thermometers are
significantly lower than the primary temperature of 5 MeV which is
shown as the dotted line in the four panels. There seems to be
a Z dependence in $T_{iso}$. $T_{iso}(^{15,16}O)$ is about 0.5 MeV
lower than $ T_{iso}(^{11,12}C)$ which is only slightly lower
($~0.2$ MeV) than $ T_{iso}(^{3,4}He)$. In addition, there is also
a trend of isotopic temperature values decreasing as a function of
$A_{1}$. The lower temperatures reflect increasing contributions
of multi-step secondary decay contributions. As these multi-step
contributions originate from the decay from an ensemble of
unstable nuclei that are less excited than the original ensemble,
it has the effect of making the system appear cooler.

For comparison, we use the corresponding isotope temperatures
extracted from the data obtained in the central collisions of
$^{124}Sn+^{124}Sn$ reactions at E/A=50 MeV \cite{txliu} shown as
solid squares(top right panel), circles (bottom left panel) and
stars (bottom right panel)\ for $T_{iso}(^{3,4}He)$, $
T_{iso}(^{11,12}C)$, $T_{iso}(^{15,16}O)$, respectively in Figure
\ref{fig:tlio}. The calculated ISMM isotopic temperatures (lines)
follow the trends of the corresponding experimental values.
Despite the fact that the parameters in the ISMM calculations have
not been optimized, the calculated temperatures of
$T_{iso}(^{11,12}C)$ and $T_{iso}(^{15,16}O)$ (bottom panels) are
nearly the same as the data within the theoretical uncertainties,
which indicates that the IMF's distributions can be well
reproduced in an appropriate equilibrium model.

However, the experimental $T_{iso}(^{3,4}He)$ temperatures (solid
squares in top right panel) are systematically higher than the
corresponding ISMM values (dot dashed line). As these thermometers
derive their sensitivity to the temperature from the large binding
energy difference between $^{3}$He and $^{4}$He, the difficulty in
reproducing these quantities may arise if there are significant
nonequilibrium production mechanisms for light particles such as
$^{3}$He \cite{xi98,xi98a}. To illustrate this effect, we assumed
that 2/3 of the measured $^{3}$He yield is of a non-thermal
origin. This increases the $^{3}$He yield by a factor of three and
the new calculations are shown as the solid line in the top right
panel. The resulting apparent temperatures are nearly the same as
the experimental data. This simple assumption explains the
discrepancies between $T_{iso}(^{3,4}He)$ and $T_{iso}(^{11,12}C)$
observed experimentally. However, the present calculations also
suggest that sequential decays have a much larger effects on
$T_{iso}(^{11,12}C)$ and  $ T_{iso}(^{15,16}O)$ than previously
assumed \cite{xi98}.

To illustrate the importance of using an accurate sequential decay
code to decay the primary hot fragments before data can be
accurately compared, Table I contains the experimental measured
isotope temperatures in the fourth column. Predicted temperatures
from the ISMM using the MSU-DECAY code are plotted in the fifth
column. As shown in Figure 16 and Table I, there is a close
correspondence in the fluctuations of the temperature between the
ISMM and observed temperatures. However, if one uses the SMM code
of Ref. [4], which contains a Fermi-break up decay mechanism for
excited fragments and utilizes schematic structure information to
calculate the secondary decays, the fluctuations in the
temperature, listed in the last column in Table I, are much larger
than those observed in the data. In this respect, one should
especially note those involving $^{8,9}$Li; $^{9,10}$Be,
$^{12,13}$B, and $^{17,18}$O where the calculated $T_{app}$ differ
from the data by more than a factor of two. The discrepancies in
the predicted ratios are significantly larger still, by a factor
of $\Delta B/T_{iso}$, according to Eq. (41).

\section{Summary}

The multifragmentation of excited nuclear systems produces excited
fragments that decay into the observed ground state nuclei by
mechanisms that are strongly influenced by the ground and excited
state spins and energies of the fragments and by their decay
branching ratios. Prior equilibrium multifragmentation models
employed approximate descriptions for these quantities that are
insufficiently accurate to describe the new isotopically resolved
data now becoming available \cite{Xu00,txliu}. In this paper, we
include this information self-consistently, building the improved
statistical multifragmentation model (ISMM) upon
the foundations of ref. \cite{smm, Sneppen87}. The main differences between
the properties of the hot systems we calculate and those
calculated in ref. \cite{smm, Sneppen87} can be attributed to the more
accurate expression for the binding energies that we employ; the
structure of the low-lying states of the fragments plays little
role in properties of the hot system. These structure effects
become critical when the fragments cool later by secondary decay.

Our calculations call many of the previous conclusions of
equilibrium multifragmentation models into question. In
particular, we have found that the SMM85 and other similar
calculations tended to overpredict the yields of heavy fragments,
and, consequently, to underestimate those of the lighter ones.
More importantly, we find that isotopic yields and observables
like the isotopic temperatures require careful attention to the
structure of the excited fragments. Thus, prior calculations of
these isotopic observables using models that do not include such
structure information accurately may be unreliable and lead to questionable
conclusions.

\acknowledgements

We would like to acknowledge the MCT/FINEP/CNPq (PRONEX) program,
under contract \#41.96.0886.00, CNPq, FAPERJ, and FUJB for partial
financial support. This work was supported in part by the National
Science Foundation under Grant No. PHY-01-10253 and INT-9908727.

\newpage
\begin{table}[ptb]
\caption{List of isotopic thermometers plotted in Figure 16.}
\label{tab:tiso}%
\begin{tabular}{|cccccc|}
IMF-meters & $\Delta B$ & a & $T_{iso}$(Data) &$T_{iso}$(ISMM) & $T_{iso}$(SMM)[20]\\ \hline
$^{6,7}$Li/$^{11,12}$C & 11.472 & 5.898 & 3.740 & 3.315 &  3.625\\
$^{7,8}$Li/$^{11,12}$C & 16.690 & 5.361 &  3.244 & 3.212 &  4.419\\
$^{8,9}$Li/$^{11,12}$C & 14.658 & 3.351 &  3.146 & 3.065 &  1.014\\
$^{9,10}$Be/$^{11,12}$C & 11.910 & 1.028 & 5.643 & 5.102 &  12.561\\
$^{11,12}$B/$^{11,12}$C & 15.352 & 3.000 & 3.651 & 3.154 &  3.928\\
$^{12,13}$B/$^{11,12}$C & 13.844 & 5.278 & 3.720 & 3.031 &  1.636\\
$^{12,13}$C/$^{11,12}$C & 13.776 & 7.917 & 3.418 & 3.078 &  3.608\\
$^{13,14}$C/$^{11,12}$C & 10.545 & 1.962 & 3.288 & 2.949 &  2.590\\
$^{15,16}$N/$^{11,12}$C & 16.233 & 9.669 & 2.767 & 2.564 &  2.716\\
$^{16,17}$O/$^{11,12}$C & 14.578 & 23.069 & 2.648 & 2.443 &  2.555\\
$^{17,18}$O/$^{11,12}$C & 10.678 & 0.637 & 6.921 & 6.009 & 4.514\\
$^{6,7}$Li/$^{15,16}$O & 8.413 & 3.050 & 2.273 & 2.352 & 2.209\\
$^{7,8}$Li/$^{15,16}$O & 13.631 & 2.773 &  2.636 &  2.565 & 3.084 \\
$^{8,9}$Li/$^{15,16}$O & 11.599 & 1.733 &  2.476 &   2.368 & 0.768 \\
$^{9,10}$Be/$^{15,16}$O & 8.851 & 0.532 &  4.143 &   3.610 & 5.562 \\
$^{11,12}$B/$^{15,16}$O & 12.293 & 1.551 &  2.906 &  2.466 & 2.701 \\
$^{12,13}$B/$^{15,16}$O & 10.785 & 2.729 &  2.109 &  2.303 & 1.184 \\
$^{12,13}$C/$^{15,16}$O & 10.717 & 4.094 &  2.643 &  2.334 & 2.402 \\
$^{13,14}$C/$^{15,16}$O & 7.486 & 1.014 &  2.316 &  2.270 & 1.588 \\
$^{15,16}$N/$^{15,16}$O & 13.174 & 5.000 &  2.236 &  2.043 & 1.990 \\
$^{16,17}$O/$^{15,16}$O & 11.519 & 11.930 &  2.083 &  1.893 & 1.814 \\
$^{17,18}$O/$^{15,16}$O & 7.619 & 0.330  &  4.863 &  4.027 & 2.523 \\
$^{6,7}$Li/$^{3,4}$He & 13.328 & 2.183 & 5.693 &  3.632 &  4.708  \\
$^{7,8}$Li/$^{3,4}$He & 18.546 & 1.984 &  4.197 &  3.431 &  5.386 \\
$^{8,9}$Li/$^{3,4}$He & 16.514 & 1.240 &  4.200 &  3.309 &  1.169 \\
$^{9,10}$Be/$^{3,4}$He & 13.766 & 0.380 &  9.938 &  5.413 &  22.410 \\
$^{11,12}$B/$^{3,4}$He & 17.208 & 1.110 &  4.948 &  3.390 &  4.814 \\
$^{12,13}$B/$^{3,4}$He & 15.700 & 1.953 &  3.599 &  3.287 &  1.931 \\
$^{12,13}$C/$^{3,4}$He & 15.632 & 2.930 &  4.731 &  3.337 &  4.487 \\
$^{13,14}$C/$^{3,4}$He & 12.401 & 0.726 &  5.000 &  3.276 &  3.319 \\
$^{15,16}$N/$^{3,4}$He & 18.089 & 3.578 &  3.519 &  2.766 &  3.206 \\
$^{16,17}$O/$^{3,4}$He & 16.434 & 8.536 &  3.439 &  2.661 & 3.059 \\
$^{17,18}$O/$^{3,4}$He & 12.534 & 0.236 &  15.334 & 6.311 & 6.170 \\
\end{tabular}
\end{table}

\begin{figure}[tbp]
\includegraphics[width=11cm]{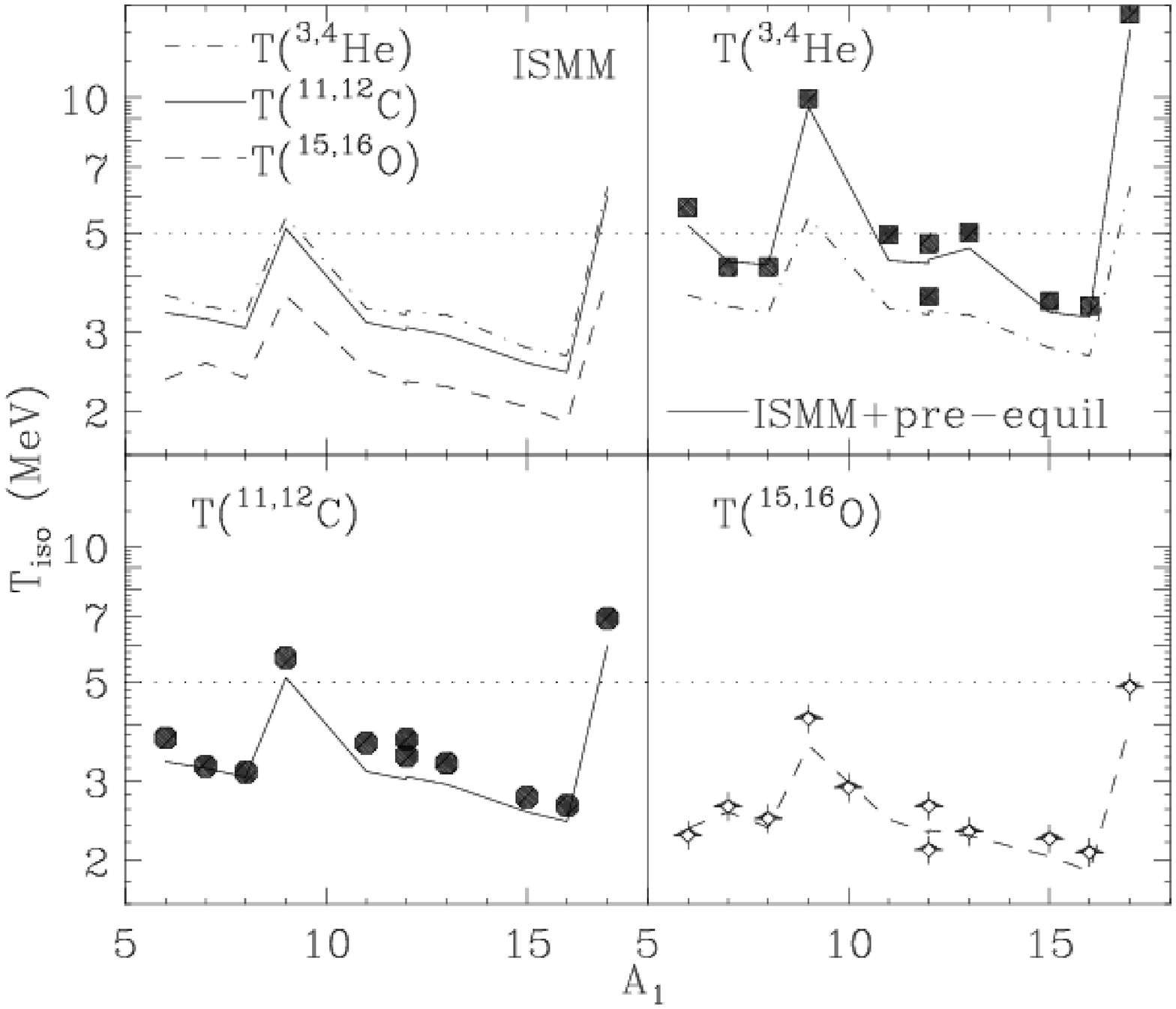}
\caption{Isotopic temperatures extracted from three types of
thermometers (see table \ref{tab:tiso}).
Experimental data are shown as the symbols. The lines are the
calculations by the ISMM. For reference, the primary temperature
of 5 MeV calculated from the ISMM is shown as the horizontal dotted lines.
For details see text.} \label{fig:tlio}
\label{fig:tlio}
\end{figure}


\begin{references}

\bibitem{Dasgupta01} S. Das Gupta, A. Z. Mekjian and M.B. Tsang,
   Adv. Nucl. Phys. \textbf{26}, 91 (2001).

\bibitem{bowman91} D.R. Bowman, G.F. Peaslee, R.T. de Souza, N. Carlin, C.K.
Gelbke, W.G. Gong, Y.D. Kim, M.A. Lisa, W.G. Lynch, L. Phair, M.B.
Tsang, C. Williams, N. Colonna, K. Hanold, M.A. McMahan, G.J.
Wozniak, L.G. Morreto, and W.A. Friedman, Phys. Rev. Lett. {\bf
67}, 1527 (1991).

\bibitem{tsang93} M.B. Tsang, W.C.Hsi, W.G. Lynch, D.R. Bowman, C.K. Gelbke,
M.A. Lisa, G.F. Peaslee, G.J. Kunde, M.L. Begemann-Blaich, T.
Hoffman, J. Hubele, J. Kempter, P. Kreutz, W.D. Kunze, V.
Lindenstruth, U. Lynen, M.Mang, W.F.J. Mueller, M. Neumann, B.
Ocker, C.A. Ogilvie, J.Pochodzalla, F. Rosenberger, H. Sann, A.
Schuettauf, V. Serfling, W. Trautmann, A. Tucholski, A. Worner, B.
Zwieglinski, G. Raciti, G. Immen, R.J. Charity, L.G. Sobotka, I.
Iori, A. Moroni, R. Scardoni, A. Ferrero, W. Seidel, L. Stuttge,
A. Cosmo, W.A. Friedman, and G. Peilert, Phys. Rev. Lett. {\bf
71}, 1502 (1993).

\bibitem{williams97} C. Williams, W. G. Lynch, C. Schwarz, M. B. Tsang, W.
C. Hsi, M. J. Huang, D. R. Bowman, J. Dinius, C. K. Gelbke, D. O.
Handzy, G. J. Kunde, M. A. Lisa, G. F. Peaslee, L. Phair, A.
Botvina, M-C. Lemaire and S. R. Souza, G. Van Buren, R. J.
Charity, and L. G. Sobotka, U. Lynen, J. Pochodzalla, H. Sann, and
W. Trautmann, D. Fox and R. T. de Souza, and N. Carlin, Phys. Rev.
{\bf C55}, R2132 (1997).

\bibitem{ogilvie91} C.A. Ogilvie, J.C. Adloff, M. Begemann-Blaich, P.
Bouissou, J. Hubele, G. Imme, P. Kreutz, G.J. Kunde, S. Leray, V.
Lindenstruth, Z. Liu, U. Lynen, R.J. Meijer, U. Milkau, W.F.J.
Muller, C. Ngo, J. Pochodzalla, G. Raciti, G. Rudolf, H. Sann, A.
Schuttauf, W. Seidel, L. Stuttge, W. Trautmann, and A. Tucholski,
Phys. Rev. Lett. {\bf 67}, 1214 (1991).

\bibitem{schuttauf96} A. Schuttauf, W.D. Kunze, A. Worner, M.
Begemann-Blaich, Th. Blaich, D.R. Bowman, R.J. Charity, A. Cosmo,
A. Ferrero, C.K. Gelbke, C. Gross, W.C. Hsi, J. Hubele, G. Imme,
I. Iori, J. Kempter, P. Kreutz, G.J. Kunde, V. Lindenstruth, M.A.
Lisa, W.G. Lynch, U. Lynen, M. Mang, T. Mohlenkamp, A. Moroni,
W.F.J. Muller, M. Neumann, B. Ocker, C.A. Ogilvie, G.F. Peaslee,
J. Pochodzalla, G. Raciti, F. Rosenberger, Th. Rubehn, H. Sann, C.
Schwarz, W. Seidel, V. Serfling, L.G. Sobotka, J. Stroth, L.
Stuttge, S. Tomasevic, W. Trautmann, A. Trzcinski, M.B. Tsang, A.
Tucholski, G. Verde, C.W. Williams, E. Zude and B. Zwieglinski,
Nucl. Phys. {\bf A607}, 457 (1996).

\bibitem{hsi97}  W.-c. Hsi, K. Kwiatkowski, G. Wang, D.S. Bracken, E.
Cornell, D.S. Ginger, V.E. Viola, N.R. Yoder, R.G. Korteling, F.
Gimeno-Nogures, E. Ramakrishnan, D. Rowland, S.J. Yennello, M.J.
Huang, W.G. Lynch, M.B. Tsang, H. Xi, Y.Y. Chu, S. Gushue, L.P.
Remsberg, K.B. Morley, and H. Breuer, Phys. Rev. Lett. {\bf 79},
617 (1997).

\bibitem{bowman93} D.R. Bowman, G.F. Peaslee, N. Carlin, R.T. de Souza, C.K.
Gelbke, W.G. Gong, Y.D. Kim, M.A. Lisa, W.C. Lynch, L. Phair, M.B.
Tsang, C. Williams, N. Colonna, K. Hanold, M.A. McMahan, G.J.
Wozniak, and L.G. Moretto, Phys. Rev. Lett. {\bf 70}, 3534 (1993)

\bibitem{popescu98} R. Popescu, T. Glasmacher, J.D. Dinius, S.J. Gaff, C.K.
Gelbke, D.O. Handzy, M.J. Huang, G.J. Kunde, W.G. Lynch, L.
Martin, C.P. Montoya, M.B. Tsang, N. Colonna, L. Celano, G.
Tagliente, G.V. Margagliotti, P.M. Milazzo, R. Rui, G. Vannini, M.
Bruno, M. D'Agostino, M.L. Fiandri, F. Gramegna, A. Ferrero, I.
Iori, A. Moroni, F. Petruzzelli, P.F. Mastinu, L. Phair, and K.
Tso, Phys. Rev. {\bf C58}, 270 (1998).

\bibitem{Beaulieu00} L. Beaulieu, T. Lefort, K. Kwiatkowski, R. T. de Souza,
W.-c. Hsi, L. Pienkowski, B. Back, D. S. Bracken, H. Breuer, E.
Cornell, F. Gimeno-Nogues, D. S. Ginger, S. Gushue, R. G.
Korteling, R. Laforest, E. Martin, K. B. Morley, E. Ramakrishnan,
L. P. Remsberg, D. Rowland, A. Ruangma, V. E. Viola, G. Wang, E.
Winchester, and S. J. Yennello, Phys. Rev. Lett. {\bf 84}, 5971
(2000).

\bibitem{Fox93} D. Fox, R.T. deSouza, L. Phair, D.R. Bowman, N. Carlin, C.K.
Gelbke, W.G. Gong, Y.D. Kim, M.A. Lisa, W.G. Lynch, G.F. Peaslee,
M.B. Tsang, and F. Zhu, Phys. Rev. {\bf C47}, R421 (1993).

\bibitem{Wang99} G. Wang, K. Kwiatkowski, D.S. Bracken, E. Renshaw Foxford,
W.-c. Hsi, K.B. Morley, V.E. Viola, N.R. Yoder, C. Volant, R.
Legrain, E.C. Pollacco, R.G. Korteling, W.A. Friedman, A. Botvina,
J. Brzychczyk, and H. Breuer, Phys. Rev. {\bf C60}, 014603 (1999).

\bibitem{Gross97} D.H.E. Gross, Phys. Rep. {\bf 279}, 119 (1997).

\bibitem{Bondorf95} J.P. Bondorf, A.S. Botvina, A.S. Iljinov, I.N.
Mishustin, K. Sneppen, Phys. Rep. {\bf 257}, 133 (1995).

\bibitem{Souza00} S.R. Souza, W.P. Tan, R. Donangelo, C.K. Gelbke, W.G.
Lynch, M.B. Tsang, Phys. Rev. {\bf C62}, 064607 (2000).

\bibitem{Lamb78} D.Q. Lamb, J.M. Lattimer, C.J. Pethick, D.G. Ravenhall,
Phys. Rev. Lett. {\bf 41}, 1623 (1978).

\bibitem{Daniel79} P. Danielewicz, Nucl. Phy. {\bf A314}, 465 (1979).

\bibitem{Jaqaman83} H. Jaqaman, A.Z. Mekjian, L. Zamick, Phys. Rev. {\bf C27}%
, 2782 (1983).

\bibitem{dagostino96} M. D'Agostino, A.S. Botvina, P.M. Milazzo, M. Bruno,
G.J. Kunde, D.R. Bowman , L. Celano, N. Colonna, J.D. Dinius, A.
Ferrero, M.L. Fiandri, C.K. Gelbke, T. Glasmacher, F. Gramegna,
D.O. Handzy, D. Horn, W.C. Hsi, M. Huang, I. Iori, M.A. Lisa, W.G.
Lynch, L. Manduci, G.V. Margagliotti, P.F. Mastinu, I.N. Mishustin
, C.P. Montoya, A. Moroni, G.F. Peaslee , F. Petruzzelli, L.
Phair, R. Rui, C. Schwarz, M.B. Tsang, G. Vannini, and C.
Williams, Phys. Lett. {\bf B371}, 175 (1996).

\bibitem{huang} M.J. Huang, H. Xi, W.G. Lynch, M.B. Tsang, J.D. Dinius, S.J. Gaff, C.K.
Gelbke, T. Glasmacher, G.J. Kunde, L. Martin, C.P. Montoya, E.
Scannapiecoet, P.M. Milazzo, M. Azzano, G.V. Margagliotti, R. Rui,
G. Vannini, N. Colonna, L. Celano, G. Tagliente, M. D'Agostino, M.
Bruno, M.L. Fiandri, F. Gramegna, A. Ferrero, I. Iori, A. Moroni,
F. Petruzzelli, P.F. Mastinu, Phys. Rev. Lett. {\bf 78}, 1648
(1997).

\bibitem{tsang97} M.B. Tsang, W.G. Lynch, H. Xi, W.A. Friedman, Phys. Rev.
Lett. {\bf 78}, 3836 (1997).

\bibitem{Xu00} H.S. Xu, M.B. Tsang, T.X. Liu, X.D. Liu, W.G. Lynch, W.P.
Tan, A. Vander Molen, G. Verde, A. Wagner, H.F. Xi, C.K. Gelbke, L.
Beaulieu, B. Davin, Y. Larochelle, T. Lefort, R.T. de Souza, R. Yanez, V.E.
Viola, R.J. Charity, L.G. Sobotka, Phys. Rev. Lett. {\bf 85}, 716 (2000).

\bibitem{reisdorf97} W. Reisdorf, D. Best, A. Gobbi, N. Herrmann, K.D.
Hildenbrand, B. Hong, S.C. Jeong, Y. Leifels, C. Pinkenburg, J.L.
Ritman, D. Schull, U. Sodan, K. Teh, G.S. Wang, J.P. Wessels, T.
Wienold, J.P. Alard, V. Amouroux, Z. Basrak, N. Bastid, I.
Belyaev, L. Berger, J. Biegansky, M. Bini, S. Boussange, A. Buta,
R. Caplar, N. Cindro, J.P. Coffin, P. Crochet, R. Dona, P.
Dupieux, M. Dzelalija, J. Ero, M. Eskef, P. Fintz, Z. Fodor, L.
Fraysse, A. Genoux-Lubain, G. Goebels, G. Guillaume, Y. Grigorian,
E. Hafele, S. Holbling, A. Houari, M. Ibnouzahir, M. Joriot, F.
Jundt, J. Kecskemeti, M. Kirejczyk, P. Koncz, Y. Korchagin, M.
Korolija, R. Kotte, C. Kuhn, D. Lambrecht, A. Lebedev, A. Lebedev,
I. Legrand, C. Maazouzi, V. Manko, T. Matulewicz, P.R. Maurenzig,
H. Merlitz, G. Mgebrishvili, J. Mosner, S. Mohren, D. Moisa, G.
Montarou, I. Montbel, P. Morel, W. Neubert, A. Olmi, G. Pasquali,
D. Pelte, M. Petrovici, G. Poggi, P. Pras, F. Rami, V. Ramillien,
C. Roy, A. Sadchikov, Z. Seres, B. Sikora, V. Simion, K.
Siwek-Wilczynska, V. Smolyankin, N. Taccetti, R. Tezkratt, L.
Tizniti, M. Trzaska, M.A. Vasiliev, P. Wagner, K. Wisniewski, D.
Wohlfarth, and A. Zhilin, Nucl. Phys. {\bf A612}, 493 (1997).

\bibitem{xi98} H.F. Xi, G.J.Kunde, O. Bjarki, C.K. Gelbke, R.C.
Lemmon, W.G. Lynch, D. Magestro, R. Popescu, R.Shomin, M.B. Tsang,
A.M. Vandermolen, G.D. Westfall G. Imme, V. Maddalena, C.
Nociforo, G. Raciti, G. Riccobene, F.P. Romano, A. Saija, C.
Sfienti, S. Fritz, C. Gro\ss , T. Odeh, C. Schwarz, A. Nadasen, D.
Sisan, K.A.G. Rao, Phys. Rev. {\bf C58}, R2636 (1998).

\bibitem{Johnston96} H. Johnston, T. White, J. Winger, D. Rowland, B. Hurst,
F. Gimeno-Nogues, D. O'Kelly S.J. Yennello, Phys. Lett. {\bf
B371}, 186 (1996).

\bibitem{tsa02} M. B. Tsang, R. Shomin, O. Bjarki, C. K. Gelbke, G. J.
Kunde, R. C. Lemmon, W. G. Lynch, D. Magestro, R. Popescu, A. M.
Vandermolen, G. Verde, G. D. Westfall, H. F. Xi, W. A. Friedman,G.
Imme, V. Maddalena, C. Nociforo, G. Raciti, G. Riccobene, F. P.
Romano, A. Saija, C. Sfienti, S. Fritz, C. Gro\ss , T. Odeh, C.
Schwarz,A. Nadasen, D. Sisan, and K. A. G. Rao, Phys. Rev. {\bf
C66}, 044618 (2002).

\bibitem{smm} J.P. Bondorf, R.Donangelo, I.N. Mishustin, C.J. Pethick, H.
Schulz, and K. Sneppen, Nucl.\ Phys. {\bf A443}, 321 (1985); {\it
ibid} {\bf A444}, 460 (1985); {\it ibid} {\bf A448}, 753 (1986).

\bibitem{Sneppen87} K.\ Sneppen, Nucl.\ Phys.\ {\bf A470}, 213 (1987).

\bibitem{Botvina87} A.S. Botvina, A.S. Iljinov, I.N. Mishustin, J.P.
Bondorf, R. Donangelo, and K. Sneppen, Nucl. Phys. {\bf A475}, 663
(1987).

\bibitem{Dasgupta98} S. Das Gupta and A. Z. Mekjian, Phys. Rev. {\bf C57}, 1361
(1998)

\bibitem{ajzenberg} F. Ajzenberg-Selove, Nucl. Phys. {\bf A460}, 1 (1986); Nucl.
Phys. {\bf A449}, 1 (1986); Nucl. Phys. {\bf A475}, 1 (1987); ;
Nucl. Phys. {\bf A490}, 1 (1988); Nucl. Phys. {\bf A506}, 1
(1990); Nucl. Phys. {\bf A523}, 1 (1991).

\bibitem{Audi95} G.\ Audi and A.\ H.\ Wapstra, Nucl.\ Phys.\ {\bf A595}, 409
(1995).

\bibitem{table} R.B. Firestone, V.S. Shirley, C.M. Baglin, S.Y.F. Chu, and
J. Zipkin, {\it Table of Isotopes}, John Wiley \& Sons, Inc., (1996);
Evaluated Nuclear Structure Data File (ENSDF), maintained by the National
Nuclear Data Center (NNDC), Brookhaven National Laboratory.

\bibitem{Hsi94} W.-c. Hsi, G.J. Kunde, J. Pochodzalla, W.G. Lynch, M.B.
Tsang, M.L. Begemann-Blaich, D.R. Bowman, R.J. Charity, A. Cosmo,
A. Ferrero, C.K. Gelbke, T. Glasmacher, T. Hofmann, G. Imme, I.
Iori, J. Hubele, J. Kempter, P. Kreutz, W.D. Kunze, V.
Lindenstruth, M.A. Lisa, U. Lynen, M. Mang, A. Moroni, W.F.J.
Mueller, N. Neumann, B. Ocker, C.A. Ogilvie, G.F. Peaslee, G.
Raciti, F. Rosenberger, H. Sann, R. Sardaoni, A. Schuettauf, C.
Schwarz, W. Seidel, V.Serfling, L.G. Sobotka, L. Stuttge, W.
Trautmann, A. Tucholski, C. Williams, A. Woerner, and B.
Zwieglinski, Phys. Rev. Lett. {\bf 73}, 3367 (1994).

\bibitem{Jeong94} S.C. Jeong, N. Herrmann, J. Randrup, J.P. Alard, Z.
Basrak, N. Bastid, I.M. Belaev, M. Bini, T. Blaich, A. Buta, R.
Caplar, C. Cerruti, N. Cindro, J.P. Coffin, R. Dona, P. Dupieux,
J. Ero, Z.G. Fan, P. Fintz, Z. Fodor, R. Freifelder, L. Fraysse,
S. Frolov, A. Gobbi, Y. Grigorian, G. Guillaume, K.D. Hildenbrand,
S. Holbling, A. Houari, M. Jorio, F. Jundt, J. Kecskemeti, P.
Koncz, Y. Korchagin, R. Kotte, M. Kramer, C. Kuhn, I. Legrand, A.
Lebedev, C. Maguire, V. Manko, T. Matulewicz, G. Mgebrishvili, J.
Mosner, D. Moisa, G. Montarou, P. Morel, W. Neuberg, A. Olmi, G.
Pasquali, D. Pelte, M. Petrovici, G. Poggi, F. Rami, W. Reisdorf,
A. Sadchikov, D. Schull, Z. Seres, B. Sikora, V. Simion, S.
Smolyankin, U. Sodan, K. Teh, R. Tezkratt, M. Trzaska, M.A.
Vasilev, P. Wagner, J.P. Wessels, T. Wienold, Z. Wilhelmi, D.
Wohlfarth, and A.V. Zhilin, Phys. Rev. Lett. {\bf 72}, 3468
(1994).

\bibitem{li93} B.-A. Li, A.R. De Angelis, and D.H.E. Gross, Phys. Lett.
{\bf B303}, 225 (1993).

\bibitem{botvina95} A.S. Botvina, I.N. Mishustin, M. Begemann-Blaich, J.
Hubele, G. Imme, I. Iori, P. Kreutz, G.J. Kunde, W.D. Kunze, V.
Lindenstruth, U. Lynen, A. Moroni, W.F.J. Muller, C.A. Ogilvie, J.
Pochodzalla, G. Raciti, Th. Rubehn, H. Sann, A. Schuttauf, W.
Seidel, W. Trautmann, and A. Worner, Nucl. Phys. {\bf A584}, 737
(1995).

\bibitem{lauret98} J. Lauret, S. Albergo, F. Bieser, N.N. Ajitanand, J.M.
Alexander, F.P. Brady, Z. Caccia, D. Cebra, A.D. Chacon, J.L.
Chance, Y. Choi, P. Chung, S. Costa, P. Danielewicz, J.B. Elliott,
M. Gilkes, J.A. Hauger, A.S. Hirsch, E.L. Hjort, A. Insolia, M.
Justice, D. Keane, J. Kintner, R.A. Lacey, V. Lindenstruth, M.A.
Lisa, H.S. Matis, R. McGrath, M. McMahan, C. McParland, W.F.J.
M\"{u}ller, D.L. Olson, M.D. Partlan, N.T. Porile, R. Potenza, G.
Rai, J. Rasmussen, H.G. Ritter, J. Romanski, J.L. Romero, G.V.
Russo, H. Sann, R. Scharenberg, A. Scott, Y. Shao, B.K.
Srivastava, T.J.M. Symons, M. Tincknell, C. Tuv\`{e}, S. Wang, P.
Warren, T. Wienold, H.H. Wieman, and K. Wolf, Phys. Rev. {\bf
C57}, R1051 (1998).

\bibitem{beaulieu00} L. Beaulieu, T. Lefort, K. Kwiatkowski, R. T. de Souza,
W.-c. Hsi, L. Pienkowski, B. Back, D. S. Bracken, H. Breuer, E.
Cornell, F. Gimeno-Nogues, D. S. Ginger, S. Gushue, R. G.
Korteling, R. Laforest, E. Martin, K. B. Morley, E. Ramakrishnan,
L. P. Remsberg, D. Rowland, A. Ruangma, V. E. Viola, G. Wang, E.
Winchester, and S. J. Yennello, Phys. Rev. Lett. {\bf 84}, 5971
(2000).


\bibitem{wigner} E. Wigner and F. Seitz, Phys. Rev. {\bf 46}, 509 (1934).

\bibitem{Tan01} W.P. Tan, B.-A. Li, R. Donangelo, C.K. Gelbke, M.-J. van
Goethem, X.D. Liu, W.G. Lynch, S. Souza, M.B. Tsang, G. Verde, A. Wagner,
H.S. Xu, Phys. Rev. {\bf C64}, 051901R (2001).

\bibitem{tsang01} M.B Tsang, C.K. Gelbke, X.D. Liu, W.G. Lynch, W.P. Tan, G.
Verde, H.S. Xu, W. A. Friedman, R. Donangelo, S. R. Souza, C.B.
Das, S. Das Gupta, D. Zhabinsky, Phys. Rev. {\bf C64}, 54615
(2001).

\bibitem{Preston} M.\ A.\ Preston and R.\ K.\ Bhaduri, {\it Structure of the
Nucleus}, Addison-Wesley publishing company, Massachusetts (1975).

\bibitem{Meyer} W.D. Myers and W.J. Swiatecki, Nucl. Phys. {\bf 81}, 1 (1966).

\bibitem{Gilbert65} A. Gilbert and A.G.W. Cameron, Can. J. Phys. {\bf 43},
1446 (1965).

\bibitem{Chen88} Z. Chen and C.K. Gelbke, Phys. Rev. {\bf C38}, 2630 (1988).

\bibitem{charity} R.J.~Charity, M.A.~McMahan, G.J.~Wozniak, R.J.~McDonald,
L.G.~Moretto, D.G.~Sarantites, L.G.~Sobotka, G.~Guarino,
A.~Panteleo, L.~Fiore, A.~Gobbi and K.~Hildenbrand, Nucl. Phys.
{\bf A483}, 371 (1988); R.J.~Charity, M.~Korolija,
D.G.~Sarantites, and L.G.~Sobotka, Phys. Rev. {\bf C56} 873
(1997); R.J.~Charity, Phys. Rev. {\bf C58} 1073 (1998).

\bibitem{Hauser52} W. Hauser and H. Feshbach, Phys. Rev. {\bf 87}, 366
(1952).

\bibitem{dasgupta} S. Das Gupta, J. Pan, I. Kvasnikova, and C. Gale, Nucl.
Phys. {\bf A621}, 897 (1997).

\bibitem{txliu} T.X. Liu et al, nucl-ex/0210004.

\bibitem{Albergo} S. Albergo, S. Costa, E. Costanzo, and A Rubbino, Nuovo
Cimento {\bf 89}, 1 (1985).

\bibitem{xi98a}H. Xi, M. J. Huang, W. G. Lynch, S. J. Gaff,    C. K. Gelbke,
T. Glasmacher, G. J. Kunde, L. Martin, C. P. Montoya, S. Pratt, M.
B. Tsang, W. A. Friedman, P. M. Milazzo, M. Azzano, G. V.
Margagliotti,  R.  Rui, G. Vannini, N. Colonna, L. Celano, and G.
Tagliente M. D'Agostino, M.  Bruno, M. L. Fiandri, F. Gramegna, A.
Ferrero, I. Iori, A. Moroni, F. Petruzzelli, and P. F. Mastinu,
Phys. Rev. {\bf C57}, R426 (1998).


\end{references}
\end{document}